\documentclass{amsart}
\usepackage{balance} % for balancing columns on the final page
\usepackage{natbib}
\usepackage{xspace}
\usepackage{xcolor}
\usepackage{amsmath}
\usepackage{amssymb}
\usepackage{hyperref}
\usepackage[T1]{fontenc}

\usepackage{tikz}
\usetikzlibrary{arrows.meta,calc,positioning,backgrounds,patterns,decorations,decorations.pathreplacing}
\usepackage{subcaption}
\usepackage[a4paper, margin=1.4in]{geometry}%%%%%%%%%%%%%%%%%%%%%%%%%%%%%%%%%%%%%%%%%%%%%%%%%%%%%%%%%%%%%%%%%%%%%%%%

\newcommand{\BibTeX}{\rm B\kern-.05em{\sc i\kern-.025em b}\kern-.08em\TeX}

\newcommand{\N}{\mathbb{N}}
\newcommand{\Z}{\mathbb{Z}}
\newcommand{\Q}{\mathbb{Q}}

\newcommand{\abs}[1]{\lvert #1 \rvert}
\newcommand{\parentheses}[1]{\left( #1 \right)}
\newcommand{\brackets}[1]{\left[ #1 \right]}
\newcommand{\braces}[1]{\left\{ #1 \right\}}

\newcommand{\Election}{\mathcal{E}}
\newcommand{\CandidateSet}{M}
\newcommand{\VoterSet}{N}
\newcommand{\ApprovalProfile}{A}
\newcommand{\CommitteeSize}{k}
\newcommand{\Committee}{C}
\newcommand{\CommitteePrime}{C'}
\newcommand{\ApprovalBallot}[1]{A_{#1}}
\newcommand{\CommitteeFamily}{\mathcal{C}}

\makeatletter
\newcommand{\DeclareFlexOp}[2]{%
  \expandafter\DeclareRobustCommand\expandafter#1\expandafter{%
    \@ifnextchar\bgroup{\FlexOp@with{#2}}{\ensuremath{\operatorname{#2}}\xspace}%
  }%
}
\newcommand{\FlexOp@with}[2]{%
  \ensuremath{\operatorname{#1}\!\left(#2\right)}%
}
\makeatother
\DeclareFlexOp{\pairs}{PAIRS}
\DeclareFlexOp{\cons}{CONS}
\DeclareFlexOp{\cheeger}{CHGR}

\NewDocumentCommand{\PreferencesGraph}{o}{%
  \IfNoValueTF{#1}{G}{G^{#1}}%
}
\NewDocumentCommand{\PreferencesEdges}{o}{%
  \IfNoValueTF{#1}{E}{E^{#1}}%
}
\NewDocumentCommand{\CommitteeGraph}{o m}{%
  \IfNoValueTF{#1}{G_{#2}}{G^{#1}_{#2}}%
}
\NewDocumentCommand{\CommitteeEdges}{o m}{%
  \IfNoValueTF{#1}{E_{#2}}{E^{#1}_{#2}}%
}
\NewDocumentCommand{\ABedges}{o m m}{%
  \IfNoValueTF{#1}{e(#2,#3)}{e_{#1}(#2,#3)}%
}
\NewDocumentCommand{\Neighbors}{o m}{%
  \IfNoValueTF{#1}{N\!\left(#2\right)}{N_{#1}\!\left(#2\right)}%
}
\NewDocumentCommand{\CandidateSupport}{m}{V_{#1}}

\NewDocumentCommand{\Boundary}{o m}{%
  \IfNoValueTF{#1}{\partial(#2)}{\partial_{#1}(#2)}%
}
\NewDocumentCommand{\Expansion}{o m}{%
  \IfNoValueTF{#1}{\phi(#2)}{\phi_{#1}(#2)}%
}
\newcommand{\Cheeger}[1]{h\!\left(#1\right)}
\newcommand{\Threshold}{\rho}
\newcommand{\dsmodulus}{L}
\newcommand{\dsk}{q}

\NewDocumentCommand{\vtype}{o m}{%
  \IfNoValueTF{#1}{\tau(#2)}{\tau_{#1}(#2)}%
}
\NewDocumentCommand{\TypeSet}{o m}{%
  \IfNoValueTF{#1}{\tau(#2)}{\tau_{#1}(#2)}%
}
\NewDocumentCommand{\TypeClass}{o m}{%
  \IfNoValueTF{#1}{V_{#2}}{V^{#1}_{#2}}%
}
\NewDocumentCommand{\TypeWeight}{o m}{%
  \IfNoValueTF{#1}{n_{#2}}{n^{#1}_{#2}}%
}
\newcommand{\TypeGraph}[1]{\Gamma_{\tau}(#1)}
\newcommand{\TypeGraphEdges}[1]{E_{\tau}(#1)}

\NewDocumentCommand{\Volume}{o m}{%
  \IfNoValueTF{#1}{\operatorname{vol}(#2)}{\operatorname{vol}_{#1}(#2)}%
}

\newcommand{\CIDomain}{\ensuremath{\mathrm{CI}}}
\newcommand{\VCIDomain}{\ensuremath{\mathrm{VCI}}}
\newcommand{\VIDomain}{\ensuremath{\mathrm{VI}}}

\NewDocumentCommand{\DiffusionMatrix}{o m}{%
  \IfNoValueTF{#1}{P_{#2}}{P_{#2}^{(#1)}}%
}
\newcommand{\Laplacian}[1]{L_{#1}}
\NewDocumentCommand{\SetConductance}{o m m}{%
  \IfNoValueTF{#1}{\Phi_{#2}\!\left(#3\right)}{\Phi_{#2}^{(#1)}\!\left(#3\right)}%
}
\newcommand{\ChainConductance}[1]{\Phi\!\left(#1\right)}

\def\Comments{0} % change to 0 to hide comments
\newcommand{\kibitz}[2]{\ifnum\Comments=1{\color{#1}{#2}}\fi}
\newcommand{\rmr}[1]{\kibitz{red}{[RESHEF:#1]}}
\newcommand{\gbs}[1]{\kibitz{blue}{[GILI:#1]}}

\newcommand{\todo}[1]{\kibitz{green}{[TODO:#1]}}

\newtheorem{theorem}{Theorem}
\newtheorem{lemma}[theorem]{Lemma}
\newtheorem{proposition}[theorem]{Proposition}
\newtheorem{corollary}[theorem]{Corollary}

\newtheorem{observation}{Observation}
\newtheorem{example}{Example}
\newtheorem{definition}{Definition}

\newcommand{\interval}[4]{%
  \draw[#4] (#1,#3) -- (#2,#3);
  \draw[#4] (#1,#3-0.10) -- (#1,#3+0.10);
  \draw[#4] (#2,#3-0.10) -- (#2,#3+0.10);
}

\newcommand{\VIHSourceBins}{s}
\newcommand{\VIHSourceItems}{q}
\newcommand{\VIHSourceWeight}[1]{a_{#1}}
\newcommand{\VIHSourceTarget}{B}
\newcommand{\VIHTotalWeight}{W}

\newcommand{\VIHNumPhases}{r}
\newcommand{\VIHNumPairs}{\nu}
\newcommand{\VIHMergeScale}{J}
\newcommand{\VIHPerturbBudget}{\Xi}
\newcommand{\VIHSlack}{\varepsilon}
\newcommand{\VIHItemScale}{v}
\newcommand{\VIHBranchGap}{\Lambda}
\newcommand{\VIHScaffoldScale}{H}
\newcommand{\VIHPhaseSpan}{\Omega}
\newcommand{\VIHNeutralArea}{L}
\newcommand{\VIHBoundaryTarget}{T}
\newcommand{\VIHDenominator}{Q}

\newcommand{\VIHPhaseSign}[1]{\sigma_{#1}}
\newcommand{\VIHPhaseTarget}[1]{b_{#1}}
\newcommand{\VIHCalibration}[1]{\delta_{#1}}
\newcommand{\VIHChoice}[2]{z_{#1,#2}}
\newcommand{\VIHEncodedWeight}[1]{w_{#1}}

\newcommand{\VIHPhaseOrigin}[1]{X_{#1}}
\newcommand{\VIHItemOrigin}[1]{\xi_{#1}}
\newcommand{\VIHPrefix}[1]{P_{#1}}

\newcommand{\VIHActiveEnd}{\eta}

\newcommand{\VIHScaffoldCandidate}[1]{c_{#1}}

\newcommand{\VIHScaffoldAnchor}{g_0}
\newcommand{\VIHBranchA}[2]{A_{#1,#2}}
\newcommand{\VIHBranchB}[2]{B_{#1,#2}}
\newcommand{\VIHMergeCandidate}[2]{e_{#1,#2}}
\newcommand{\VIHItemAnchor}[2]{g_{#1,#2}}

\newcommand{\VIHConstructedElection}{\Election^{\star}}

\newcommand{\VIHRightReach}[2]{R_{#1}(#2)}
\newcommand{\VIHNeutralRow}[2]{f_{#1}(#2)}
\newcommand{\VIHNeutralPrimitive}[2]{F_{#1}(#2)}

\newcommand{\VIHBranchALabel}{A}
\newcommand{\VIHBranchBLabel}{B}

\newcommand{\newpar}[1]{\vspace{0mm}\paragraph{#1}}
\newcommand{\VICheegerExact}{\textsc{VI-Cheeger-Exact}}
\newcommand{\StrictUnaryThreePartition}{\textsc{Unary-3-Partition}}
\newcommand{\VIHSourceTriple}[1]{\mathcal{T}_{#1}}

\newenvironment{computationalproblem}[1]{%
  \par\medskip
  \noindent{\normalfont\scshape #1}\par
  \begin{list}{}{%
    \setlength{\leftmargin}{1.6cm}%
    \setlength{\labelwidth}{1.35cm}%
    \setlength{\labelsep}{0.25cm}%
    \setlength{\itemsep}{0.3ex}%
    \setlength{\parsep}{0pt}%
    \setlength{\topsep}{0.4ex}%
  }%
}{%
  \end{list}
  \medskip
}

\begin{document}

\title{Interlacing Committees with Edge-Expansion}

%%% Provide names, affiliations, and email addresses for all authors.

\author{Gili Bielous}
\address{Technion- Israel Institute of Technology}
\email{gili.bielous@campus.technion.ac.il}

\author{Reshef Meir}
  \address{Technion- Israel Institute of Technology}
\email{reshefm@technion.ac.il}

%%% Use this environment to specify a short abstract for your paper.

\begin{abstract}
Approval-based committee selection has recently been studied through the lens of \emph{interlacing}, a framework for studying polarization in approval elections, in which committees induce a similarity graph over voters. The quality of committees is then evaluated through the connectivity of the resulting graph.
We revisit the connectivity measures used in this framework and propose \emph{edge-expansion}, a classical graph-theoretic measure with well-established connections to polarization in networks, as the optimization objective. We provide a theoretical foundation for this proposal by relating interlacing to classical models of information diffusion and by comparing edge-expansion with previously proposed connectivity objectives, identifying qualitative differences in the structural properties they capture.
Finally, we study the computational complexity of optimizing edge-expansion, showing that hardness persists even for voter-interval preferences, while the problem is polynomial-time solvable for every fixed committee size, with an improved dependence on the committee size for voter-candidate-interval preferences.\rmr{make sure to update according to results}\gbs{V}
 \rmr{maybe say edge-expansion in title?}\gbs{V}
\end{abstract}
%%% Use this command to specify a few keywords describing your work.
%%% Keywords should be separated by commas.

\keywords{Computational Social Choice, Approval-Based Committee Voting, Polarization, Edge Expansion, Opinion Dynamics \rmr{really?}\gbs{I had a todo for this I swear}}
%%%%%%%%%%%%%%%%%%%%%%%%%%%%%%%%%%%%%%%%%%%%%%%%%%%%%%%%%%%%%%%%%%%

%%% The following commands remove the headers in your paper. For final 
%%% papers, these will be inserted during the pagination process.

%\pagestyle{fancy}
%\fancyhead{}

%%% The next command prints the information defined in the preamble.

\maketitle

%%%%%%%%%%%%%%%%%%%%%%%%%%%%%%%%%%%%%%%%%%%%%%%%%%%%%%%%%%%%%%%%%%%%%%%%

\section{Introduction}

In recent years, rising political polarization has become an increasingly prominent concern for democratic societies, with consequences for social cohesion, public deliberation, and the ability to reach collective decisions. At the same time, parliaments and other elected bodies have become increasingly divided and antagonistic~\cite{mehlhaff2024group}.\rmr{reference?} A natural interpretation is that this political division simply reflects polarization in the electorate: a more polarized society produces a more polarized political environment and, ultimately, more polarized representatives. However, several studies suggest that these two phenomena need not coincide~\cite{fiorina2017unstable,fiorina2011culture}. \rmr{Check again what they say, especially \cite{abramowitz2008polarization}}\gbs{you were right}In particular, this literature distinguishes between \emph{mass polarization}, occurring within the population, and \emph{elite polarization}, occurring among the elected representatives of that population~\cite{mccarty2005polarized}. From this perspective, a parliament may not merely reflect the divisions present in society. Rather, it can act as a prism through which the society is perceived, potentially amplifying weak  obscuring underlying divisions. This raises the possibility that selecting a less polarized body of representatives may reduce the  division within the electorate.

A recent paper by~\citet{dong2025selecting} \rmr{the reference number should appear} tackles this issue by proposing \emph{interlacing}, a novel model in which the population is represented by a hypergraph, with individual voters as vertices and candidates as hyperedges linking their approving voters. \rmr{you are still describing Dong work so why this sentence?}\gbs{rephrased}\rmr{not exactly equivalent but you can say it induces a voter graph}\gbs{V} This hypergraph induces a voter graph in which two voters are connected whenever they jointly approve a candidate. Each committee induces a voter network through shared representation by the selected candidates, yielding a projection of the society through the lens of the committee. Committees are then evaluated by how well the selected candidates \emph{interlace}, or interconnect, the voters. To this end, Dong et al. consider two measures. The \pairs objective measures direct pairwise voter connections via the number of edges in the resulting graph, whereas \cons captures global connectivity by counting path connected-pairs of voters.

The interlacing model rests on the intuition that greater connectivity of the induced network corresponds to greater cohesion, and thus less polarization. However, we believe that the objectives proposed by~\citet{dong2025selecting} do not capture the full structural picture of polarization. The local nature of \pairs ignores global connectivity, while \cons assigns the same score to all committees whose induced graphs are connected. Thus, $\cons$ cannot distinguish, for example, between a clique and two cohesive groups joined by a single fragile bridge. This lack of granularity is already noted by \citeauthor{dong2025selecting}, who point out settings in which \(\cons\) becomes uninformative and suggest shortest-path length as a possible refinement.

More broadly, the interlacing graph records positive relations induced by shared approval, but does not capture how disagreement between different parts of the electorate may persist or dissipate.

In this paper, we retain this voter-network perspective but analyze the robustness of the connections between different parts of the electorate using \emph{edge-expansion}. We additionally give the induced network a complementary information-diffusion interpretation, in which it governs how voters' opinions evolve. \rmr{check}\gbs{Changed it a bit, also adding to the gap narrative.}

\rmr{very good background - gap - fill the gap narrative}\gbs{see above.}

\subsection{Related work}
\paragraph{Graph measures of connectivity}
Network science characterizes community structure through cohesive groups with relatively sparse connections between them~\cite{girvan2002community}, while work on network polarization adopts closely related boundary-based notions~\cite{guerra2013measure,interian2023network}. From this perspective, the relevant question is not only how well connected the network is, but how easily a large subset can be separated from the collective. We therefore propose  \rmr{I'm not sure what each of these papers show. If all you need is from the book, then enough to cite the book}\gbs{different studies different names same graph measure} the use of \emph{edge-expansion}~\cite{alon1997edge}, also known as the \emph{isoperimetric number}~\cite{mohar1989isoperimetric} or \emph{Cheeger constant}, that measures the worst case ratio between the boundary of a group and its size. We refer to the graph parameter as the \emph{Cheeger constant}, and to the corresponding committee-selection objective as \cheeger. Intuitively, edge-expansion measures the extent to which a graph is bottlenecked, by identifying weak ties of large clusters. It is a classic graph-theoretic measure with established connections to graph clustering and community detection~\cite{buhler2009spectral} and other boundary structures used to quantify polarization and political-network separation~\cite{guerra2013measure,flamino2023political}.

\newpar{Polarization}
Polarization has received rapidly growing attention in recent years in the social choice literature, with most prior work focusing on ordinal preferences. \citet{can2015measuring} study how polarization is manifested in preference profiles, and more recent work by~\citet{faliszewski2023diversity} develops computational measures of polarization. A separate line of work considers polarizing individual alternatives. \citet{gehrlein2005probabilities} studies candidates that tend to be ranked at the extremes of voters' preferences; building on this notion, \citet{barbaro2025election} studies how different voting rules fare in avoiding the election of polarizing candidates. In multiwinner settings, \citet{delemazure2024selecting} study the selection of maximally conflicting pairs of candidates under ordinal preferences. To the best of our knowledge, \citet{dong2025selecting} were the first to study polarization in approval-based committee selection, through the interlacing framework that we build upon here.

\rmr{see if relevant: \url{https://arxiv.org/abs/2605.14983}, and  McCarty, Poole \& Rosenthal, Polarized America}\gbs{The first is probably relevant but I found out about it today so I don't want to open this can of worms if I can avoid it, it's a month old.}

\subsection{Our contribution}
Our results show that \cheeger, beyond its intuitive interpretation, substantially refines the $\pairs$ and $\cons$ measures. In particular, we show that since $\pairs$  prioritize dense local connections over bridges, it may fail to select even near-perfect committees. On unrestricted profiles, two $\cons$-optimal committees can have \cheeger scores differing by a quadratic factor in the number of voters. % CI-RESULT-SENTENCE
On \CIDomain{} profiles the separation  asymptotically linear.

A second complementary justification for using \cheeger in the context of electing committees, comes from information diffusion. Joint approval is an observable signal of similarity in electoral preferences, so the committee-induced preferences graph can be viewed as a stylized network of potential interaction or information flow. We use the classical DeGroot model~\cite{degroot1974reaching} of opinion diffusion as an additional interpretation of the interlacing model, thereby identifying polarized committees with a slower time to reach consensus among voters in the induced voter graph. The latter, in turn, can be bounded by its  \cheeger value.
%The model also allows us to introduce disagreement explicitly by adding voters' opinions. In this sense, it offers a dual perspective in which \cheeger acts as a bridge: structurally, it measures the robustness of the connections between different parts of the electorate, while in the diffusion model it reflects information bottlenecks that hinder reaching consensus efficiently. We formally derive conditions for reaching a consensus and bounds on the time to convergence in terms of optimal \cheeger value. Thus, we further distinguish between the different measures through their implications for information diffusion.

We then turn to the computational aspects of finding an optimal \cheeger committee.\rmr{Cheeger,\cheeger, or $\cheeger$ ?}\gbs{see above, Cheeger for the graph measure \cheeger for the objective ans $\cheeger$ is legacy I removed it} As is the case with \pairs and \cons, finding a \cheeger-optimal committee is NP-hard. In contrast, even  finding the \cheeger value of a fixed committee is  hard, and hardness of finding an optimal committee extends to the restricted $\VIDomain$ preferences domain. Yet, we manage to obtain a  polynomial-time algorithm for fixed $\CommitteeSize$, and quasi-polynomial complexity for polylogarithmic $\CommitteeSize$ in restricted preference domains.\rmr{why quasi?}\gbs{m choose k (committee enumeration) is quasi for k=O(log m)}

%Together, these results position \cheeger as a refinement of interlacing that captures the robustness of voter connections while also admitting a complementary information-diffusion interpretation, and delineate both the structural gains it provides and the computational cost of optimizing it.

%%%%%%%%%%%%%%%%%%%%%%%%%%%%%%%%%%%%%%%%%%%%%%%%%%%%%%%%%%%%%%%%%%%%%%%%

\section{Preliminaries}

\paragraph{Approval-Based Multiwinner Voting}
In \emph{Approval-Based Multiwinner Voting}
(see \cite{lackner2023multi} for a complete survey),
a set $\VoterSet$ of $n \in \N$ voters wish to select a committee of $\CommitteeSize \in \N$ members from a set $\CandidateSet$ of $m \in \N$ candidates.
Voter $i \in \VoterSet$ \emph{approves} a subset of candidates $\ApprovalBallot{i} \subseteq \CandidateSet$. The \emph{support} of candidate $c\in M$ is the set of voters approving $c$ given by $\CandidateSupport{c} = \braces{i\in N | c \in A_i}$. The \emph{profile} of voters $\ApprovalProfile = \parentheses{\ApprovalBallot{i}}_{i\in\VoterSet}$ defines the joint preferences of all voters.
An \emph{election} is defined by the tuple $\Election=\parentheses{\CandidateSet,\VoterSet,\ApprovalProfile,\CommitteeSize}$, i.e., the set of candidates, the set of voters, their preferences, and the target committee size.
For an election $\parentheses{\CandidateSet,\VoterSet,\ApprovalProfile,\CommitteeSize}$, the set of possible results, i.e., $\CommitteeSize$-sized subsets of $\CandidateSet$, is denoted by $\CommitteeFamily = \braces{C\subseteq \CandidateSet | \abs{C}=\CommitteeSize}$.
The result of an election is determined by a \emph{voting rule}, a function that maps elections to a committee $\Committee\in\CommitteeFamily$.

% \newpar{Chamberlin-Courant Score.}
% In approval-based multiwinner voting, it is common to measure how well an elected committee represents the population with the Chamberlin-Courant score~\cite{chamberlin1983representative,lackner2018consistent}- the number of voters who approve at least one member in the committee:
% $$
% CC\parentheses{\Committee} = \abs{\braces{i\in \VoterSet\mid \ApprovalBallot{i}\cap \Committee \ne \emptyset}}
% $$

\subsection{Interlacing Committees}
Dong et al.~\cite{dong2025selecting} formulate an approval profile as a hypergraph in which voters are vertices and each candidate induces a hyperedge containing its approving voters. We use an equivalent voter-graph representation, which is more convenient for the graph-theoretic analysis developed here.\rmr{is it really equivalent (maintains al information) or equivalent for our needs?}\gbs{I argue it is equivalent.}

\newpar{Preferences Graph.}
For an election $\Election=\parentheses{\CandidateSet,\VoterSet,\ApprovalProfile,\CommitteeSize}$, we define the \emph{preferences graph} of $\Election$ to be the unweighted undirected graph $\PreferencesGraph[\Election]=\parentheses{\VoterSet,\PreferencesEdges[\Election]}$, where
\begin{equation*}
    \PreferencesEdges[\Election]
    =
    \braces{\braces{i,j}\in\binom{\VoterSet}{2}\mid
    \ApprovalBallot{i}\cap\ApprovalBallot{j}\ne\emptyset}.
\end{equation*}

\begin{figure}[t]
    \centering

    \begin{subfigure}[b]{0.23\linewidth}
        \centering
        \begin{tikzpicture}[
            scale=.55,
            every node/.style={circle,draw,inner sep=1.5pt}
        ]
            \node (v1) at (0,1.5) {$v_1$};
            \node (v2) at (-1,0) {$v_2$};
            \node (v3) at (1,0) {$v_3$};
            \node (v4) at (2,1.5) {$v_4$};

            \draw (v1)-- node[left,draw=none] {$a$} (v2);
            \draw (v1)-- node[right,draw=none] {$b$} (v3);
            \draw (v1)-- node[above,draw=none] {$c$} (v4);
            \draw (v2)-- node[below,draw=none] {$d$} (v3);
        \end{tikzpicture}
        \caption{$\PreferencesGraph[\Election]$}
        \label{fig:running-preferences}
    \end{subfigure}
    \hfill
    \begin{subfigure}[b]{0.23\linewidth}
        \centering
        \begin{tikzpicture}[
            scale=.55,
            every node/.style={circle,draw,inner sep=1.5pt}
        ]
            \node (v1) at (0,1.5) {$v_1$};
            \node (v2) at (-1,0) {$v_2$};
            \node (v3) at (1,0) {$v_3$};
            \node (v4) at (2,1.5) {$v_4$};

            \draw (v1)-- node[left,draw=none] {$a$} (v2);
            \draw (v1)-- node[right,draw=none] {$b$} (v3);
            \draw (v2)-- node[below,draw=none] {$d$} (v3);
            \draw[dashed] (v1)-- node[above,draw=none] {$c$} (v4);
        \end{tikzpicture}
        \caption{$\Committee_0=\braces{a,b,d}$.}
        \label{fig:running-disconnected}
    \end{subfigure}
    \hfill
    \begin{subfigure}[b]{0.23\linewidth}
        \centering
        \begin{tikzpicture}[
            scale=.55,
            every node/.style={circle,draw,inner sep=1.5pt}
        ]
            \node (v1) at (0,1.5) {$v_1$};
            \node (v2) at (-1,0) {$v_2$};
            \node (v3) at (1,0) {$v_3$};
            \node (v4) at (2,1.5) {$v_4$};

            \draw (v3)-- node[below,draw=none] {$d$} (v2);
            \draw (v2)-- node[left,draw=none] {$a$} (v1);
            \draw (v1)-- node[above,draw=none] {$c$} (v4);
        \end{tikzpicture}
        \caption{$\Committee_1=\braces{a,c,d}$.}
        \label{fig:running-path}
    \end{subfigure}
    \hfill
    \begin{subfigure}[b]{0.23\linewidth}
        \centering
        \begin{tikzpicture}[
            scale=.55,
            every node/.style={circle,draw,inner sep=1.5pt}
        ]
            \node (v1) at (0,0) {$v_1$};
            \node (v2) at (-1,1.3) {$v_2$};
            \node (v3) at (0,2) {$v_3$};
            \node (v4) at (1,1.3) {$v_4$};

            \draw (v1)-- node[left,draw=none] {$a$} (v2);
            \draw (v1)-- node[left,draw=none] {$b$} (v3);
            \draw (v1)-- node[right,draw=none] {$c$} (v4);
        \end{tikzpicture}
        \caption{$\Committee_2=\braces{a,b,c}$.}
        \label{fig:running-star}
    \end{subfigure}

    \caption{The running example. (a) shows the full preferences
    graph, with each edge labeled by the candidate witnessing it.
    Panels (b)--(d) show committee-induced graphs. Dashed edge in
    panel (b) belongs to the preferences graph and absent from the
    committee graph.}
    \label{fig:running-example}\vspace{-0.0em}
\end{figure}
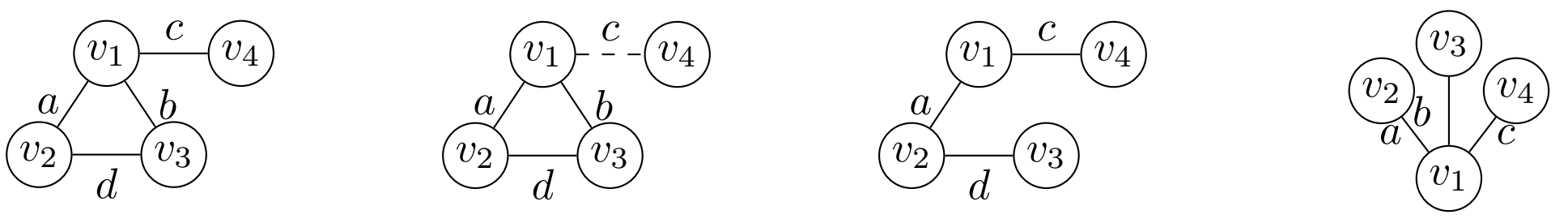

\begin{example}[Running example]
\label{ex:running}
Consider an election with voters $\VoterSet=\braces{v_1,v_2,v_3,v_4}$,
candidates $\CandidateSet=\braces{a,b,c,d}$, committee size
$\CommitteeSize=3$, and approval ballots $\ApprovalBallot{v_1}=\braces{a,b,c},
    \ApprovalBallot{v_2}=\braces{a,d},
    \ApprovalBallot{v_3}=\braces{b,d},
    \ApprovalBallot{v_4}=\braces{c}$.
For example,
$\ApprovalBallot{v_1}\cap\ApprovalBallot{v_2}=\braces{a}$,
so $\braces{v_1,v_2}$ is an edge of the preferences graph. In contrast,
$\ApprovalBallot{v_2}\cap\ApprovalBallot{v_4}=\emptyset$, so
$\braces{v_2,v_4}$ is not an edge. Thus, $\PreferencesEdges[\Election]
    =
    \braces{
        \braces{v_1,v_2},
        \braces{v_1,v_3},
        \braces{v_1,v_4},
        \braces{v_2,v_3}
    }$.
The resulting preferences graph is shown in
Figure~\ref{fig:running-example}\subref{fig:running-preferences}.
\end{example}

\rmr{must have at leat one figure of an example, maybe 2}\gbs{We now have running example with figure}
\newpar{Committee-Induced Preferences Graph.} (referred to as committee graphs for short).
For each committee $\Committee\in\CommitteeFamily$, we denote by
$\CommitteeGraph[\Election]{\Committee}=\parentheses{\VoterSet,\CommitteeEdges[\Election]{\Committee}}$
the spanning subgraph of $\PreferencesGraph[\Election]$ in which two voters are adjacent iff they jointly approve a candidate in $\Committee$:
\begin{equation*}
    \CommitteeEdges[\Election]{\Committee}
    =
    \braces{\braces{i,j}\in\PreferencesEdges[\Election]\mid
    \ApprovalBallot{i}\cap\ApprovalBallot{j}\cap\Committee\ne\emptyset}.
\end{equation*}
The election parameter is omitted from graph notation when clear from context.

\newpar{Running example (continued).}
Consider the committee $\Committee_0=\braces{a,b,d}$.
The edges $\braces{v_1,v_2}$, $\braces{v_1,v_3}$, and
$\braces{v_2,v_3}$ remain in the committee graph, witnessed respectively
by $a$, $b$, and $d$. However, although
$\braces{v_1,v_4}\in\PreferencesEdges[\Election]$, its only witnessing
candidate is $c\notin\Committee_0$. Hence this edge does not belong to
$\CommitteeEdges[\Election]{\Committee_0}$.
Consequently,
$\CommitteeGraph[\Election]{\Committee_0}$ consists of a triangle on
$v_1,v_2,v_3$ and the isolated voter $v_4$, as shown in
Figure~\ref{fig:running-example}\subref{fig:running-disconnected}.

\newpar{Connectivity Objectives.}
Following Dong et al., \pairs and \cons are numerical scores of committees. For $\Committee\in\CommitteeFamily$, define
\begin{align*}
    \pairs{\Committee,\Election}
    &:=\abs{\CommitteeEdges[\Election]{\Committee}},\\
    \cons{\Committee,\Election}
    &:=\abs{\braces{\braces{u,v}\in\binom{\VoterSet}{2}\mid u\sim_{\Committee}v}},
\end{align*}
where $u\sim_{\Committee}v$ means that $u$ and $v$ are connected by a path in $\CommitteeGraph[\Election]{\Committee}$.
We overload the objective names for the associated voting rules: $\pairs{\Election}$ and $\cons{\Election}$ denote a score-maximizing committee, subject to some tie-breaking mechanism.

An optimizing committee need not be unique; unless stated otherwise, the rule returns an arbitrary maximizer. Thus, statements about what a rule may select versus what every optimizer satisfies are distinguished explicitly.

\begin{definition}[Distinguishing Committees]
    For any committee score $f(\Committee,\Election)$, we say that $f$ \emph{distinguishes} between $\Committee,\CommitteePrime$ if their scores differ, and \emph{does not distinguish} between them otherwise.
\end{definition}

\begin{definition}[Connecting Committees]
    A committee $\Committee$ is \emph{connecting} if $\CommitteeGraph[\Election]{\Committee}$ is connected, and \emph{non-connecting} otherwise.
\end{definition}\rmr{use Definition environment}\gbs{done}

\newpar{Running example (continued).}
Besides $\Committee_0$, consider $\Committee_1=\braces{a,c,d}$.
Its committee graph is the path
$v_3-v_2-v_1-v_4$; see
Figure~\ref{fig:running-example}\subref{fig:running-path}.

Each of $\Committee_0,\Committee_1$ induce three edges, and hence $\pairs{\Committee_0,\Election}
    =
    \pairs{\Committee_1,\Election}
    =3$.
Thus, $\pairs$ does not distinguish between them and may in particular
select the non-connecting committee $\Committee_0$.

For $\cons$, the component of $\Committee_0$ containing
$v_1,v_2,v_3$ contributes all $3$ connected voter pairs,
while $v_4$ is isolated. Hence $\cons{\Committee_0,\Election}=3$.
By contrast, $\Committee_1$ is connected, so every pair of voters is
connected and $\cons{\Committee_1,\Election}
    =
    \binom{4}{2}=6$.
This is the maximum possible $\cons$ value, and therefore
$\Committee_1$ is $\cons$-optimal.

\subsection{Expansion and CHEEGER}

\paragraph{Edge-Expansion.} \rmr{source? textbook?}\gbs{done}
We use the standard edge-expansion terminology and definitions of ~\citet{mohar1989isoperimetric}, adapted to our notation.
For a graph $G=\parentheses{V,E}$ and a set $S\subseteq V$, the \emph{boundary} of $S$ is
\[
\Boundary[G]{S}=\braces{\braces{u,v}\in E\mid u\in S,\ v\notin S}.
\]
For a nonempty proper set $S$, its \emph{edge-expansion} is
\[
\Expansion[G]{S}
=
\frac{\abs{\Boundary[G]{S}}}{\min\braces{\abs{S},\abs{V\setminus S}}}.
\]
The \emph{edge-expansion}, or \emph{Cheeger constant}, of $G$ is
\[
\Cheeger{G}
=
\min_{\substack{\emptyset\ne S\subseteq V\\ \abs{S}\le \abs{V}/2}}
\frac{\abs{\Boundary[G]{S}}}{\abs{S}}.
\]
Graph parameters are omitted from $\Boundary$ and $\Expansion$ when clear from context. Intuitively, edge-expansion measures the ``bottleneckedness'' of a graph by seeking a relatively large set with few edges connecting it to the rest of the graph. The complexity of calculating the Cheeger constant of a graph is known to be NP-hard~\cite{leighton1999multicommodity,garey1974some}. \rmr{ Garey-Johnson-Stockmeyer 1974 is max-cut / simple NP-complete problems, not edge-expansion. Check out Matula \& Shahrokhi, 'Sparsest cuts and bottlenecks in graphs', Discrete Applied Math 27 (1990), https://doi.org/10.1016/0166-218X(90)90124-Y; or Mohar 1989. The same citations are reused in App. B.1.}\gbs{Matula-Shahrokhi study uniform sparsest cut, not the cardinality normalized sparsest (similar to our objective but different, not the same cuts are optimal). GJS also show NP for optimal bisection and is brought as supporting citation for Leighton-Rao that use opt bisection in their NP hardness proof for our objective.}

\newpar{\cheeger Objective.}
The \cheeger score of committee $\Committee\in\CommitteeFamily$ is $\cheeger{\Committee,\Election}
:=
\Cheeger{\CommitteeGraph[\Election]{\Committee}}$.
The associated rule $\cheeger{\Election}$ returns an arbitrary maximizing committee.

\newpar{Running example (continued).}
Consider the committee $\Committee_2=\braces{a,b,c}$.
Since $\CommitteeGraph[\Election]{\Committee_0}$ is disconnected, $\cheeger{\Committee_0,\Election}=0$.

For $\Committee_1$, consider
$S=\braces{v_2,v_3}$.  The path has only the edge
$\braces{v_1,v_2}$ crossing this cut, and therefore $\Expansion[\CommitteeGraph{\Committee_1}]{S}
    =\frac{1}{2}$.
Since the graph is connected, every nontrivial cut has at least one
boundary edge, and no admissible smaller side has more than two vertices.
Thus, $\cheeger{\Committee_1,\Election}=\frac12$.
For the star induced by $\Committee_2$, every nonempty set
$S$ with $\abs{S}\le2$ has at least $\abs{S}$ boundary edges.
Consequently, $\cheeger{\Committee_2,\Election}=1$.
The only remaining size-$3$ committee, $\braces{b,c,d}$, induces
another four-vertex path and therefore also has \cheeger score $1/2$.
Hence $\Committee_2$ is the unique \cheeger-optimal committee.

Thus, on this single election, $\pairs$ fails to distinguish any of the
displayed committees, $\cons$ distinguishes connectivity but ties the
path and the star, while \cheeger further distinguishes the robustness
of their connections.

\subsection{Spatial Preferences Domains}
\label{subsec:spatial-domains}
Besides unrestricted approval profiles, we consider three standard domains of spatial one-dimensional preferences~\cite{elkind2015structure,godziszewski2021analysis,brill2025individual}.  \rmr{here too we need a figure with an example}\gbs{V}

\begin{figure}[t]
\centering
\begin{tikzpicture}[x=0.75cm,y=0.82cm,
    pt/.style={circle,fill=black,inner sep=1.6pt},
    intstyle/.style={line width=0.9pt},
    guide/.style={densely dashed,gray!75},
    lab/.style={font=\scriptsize}
]
  \draw[thin] (0,0) -- (8,0);
  \def\pv{2.35}
  \def\rv{1.15}
  \def\pc{4.15}
  \def\rc{1.55}
  \pgfmathsetmacro{\leftint}{max(\pv-\rv,\pc-\rc)}
  \pgfmathsetmacro{\rightint}{min(\pv+\rv,\pc+\rc)}
  \fill[gray!25,opacity=0.45] (\leftint,-0.95) rectangle (\rightint,0.95);
  \draw[guide] (\leftint,-1.08) -- (\leftint,1.08);
  \draw[guide] (\rightint,-1.08) -- (\rightint,1.08);
  \draw[thin] (0,0) -- (8,0);

  \interval{\pv-\rv}{\pv+\rv}{0.58}{intstyle}
  \node[pt] at (\pv,0.58) {};
  \node[lab,below=1pt] at (\pv,0.58) {$p_v$};
  \draw[<->,thin] (\pv,0.98) -- ({\pv+\rv},0.98);
  \node[lab,above] at ({\pv+0.72*\rv},0.98) {$r_v$};

  \interval{\pc-\rc}{\pc+\rc}{-0.58}{intstyle}
  \node[pt] at (\pc,-0.58) {};
  \node[lab,above=1pt] at (\pc,-0.58) {$p_c$};
  \draw[<->,thin] ({\pc-\rc},-0.98) -- (\pc,-0.98);
  \node[lab,below] at ({\pc-0.72*\rc},-0.98) {$r_c$};
\end{tikzpicture}
\caption{Schematic illustration of \VCIDomain{} domain. Voters and candidates are embedded on a common one-dimensional spectrum represented by an interval of acceptable political views. Approval is determined by interval intersection.\rmr{keep just 1}
}
\label{fig:pref-domains}
%\vspace{-1.5em}
\end{figure}
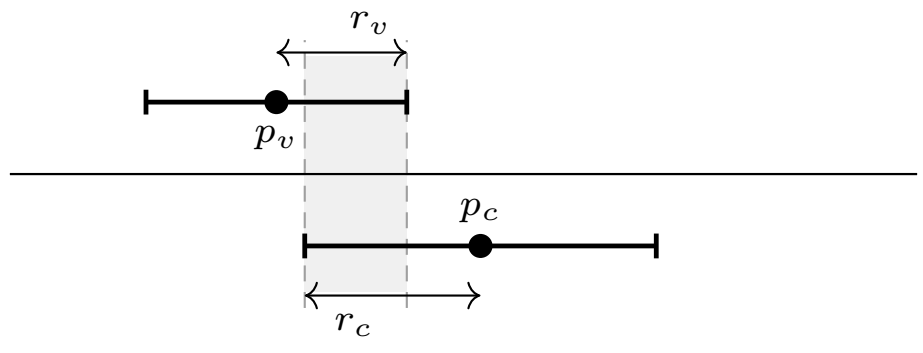

An election is in the \emph{voter-interval} (\VIDomain{}) domain if there is an ordering $v_1,\ldots,v_n$ of the voters such that every candidate support is an interval of this order. That is, for every $c\in\CandidateSet$ with $\CandidateSupport{c}\neq\emptyset$, there exist $p\le q$ such that $\CandidateSupport{c}=\braces{v_p,\ldots,v_q}$.
Similarly, an election is in the \emph{candidate-interval} (\CIDomain{}) domain if there is an ordering $c_1,\ldots,c_m$ of the candidates such that every nonempty approval ballot is an interval of this order. The \emph{voter-candidate-interval} (\VCIDomain{}) domain is a generalization of both \VIDomain{} and \CIDomain{}. An election is in the \VCIDomain{} domain if there is a set of voter positions $\braces{p_{v}}_{v\in \VoterSet}$, a set of candidate positions $\braces{p_c}_{c\in \CandidateSet}$, and a set of non-negative radii for voters and candidates $\braces{r_v}_{v\in \VoterSet}\cup\braces{r_c}_{c\in\CandidateSet}$ such that $A_i=\braces{c\in \CandidateSet | \brackets{p_{v_i}-r_{v_i},p_{v_i}+r_{v_i}}\cap\brackets{p_{c}-r_{c},p_{c}+r_{c}}\ne \emptyset}$\rmr{fixed brackets here}\gbs{:)}. Thus, voters and candidates can be represented by intervals on the real line so that a voter approves a candidate if and only if their intervals intersect. The domains are visualized schematically in Figure~\ref{fig:pref-domains}.

\newpar{Candidate Dominance.}
A candidate $c'\in\CandidateSet$ is \emph{dominated} by a candidate $c\in\CandidateSet$ if \rmr{Claude says the relation is backwards, although it seems correct to me! check anyways}\gbs{It is correct the problem was in the implication later in paragraph.} $\CandidateSupport{c'}\subsetneq\CandidateSupport{c}$.
Dong et al.~\cite{dong2025selecting} show that a dominance-free \VCIDomain{} election is \CIDomain{}. They continue to argue that in the interlacing model dominated candidates have no effect on the resulting preferences graph. Indeed, if $c'$ is dominated by $c$ then for every $i\ne j$ we have $c'\in A_i\cap A_j \rightarrow c\in A_i\cap A_j$. Moreover, replacing a dominated candidate in a committee by a candidate that dominates it weakly increases the objective value, since all objectives considered here are monotone under the addition of edges. Thus, an optimal committee can be chosen after removing dominated candidates; if this leaves fewer than $\CommitteeSize$ candidates, the committee can be completed with removed dominated candidates without changing its committee graph. Dong et al.~\cite{dong2025selecting} use the same dominance preprocessing in their polynomial-time algorithms for \pairs and \cons in the \VCIDomain{} domain.

%%%%%%%%%%%%%%%%%%%%%%%%%%%%%%%%%%%%%%%%%%%%%%%%%%%%%%%%%%%%%%%%%%%%%%%%

\section{Structural Properties}

We now compare the structural behavior of the three interlacing objectives. We first show that optimizing direct overlap can lead to severe failures of global interlacing, and then compare \cons and \cheeger once connectivity is achieved.

\subsection{PAIRS as suboptimal interlacing objective}

\begin{theorem}
\label{thm:pairs_vanishing_component}
For every $\delta>0$, there exists an election admitting a connecting committee such that the unique \pairs-optimal committee has no connected component containing more than a $\delta$-fraction of the voters. Moreover, the unique \pairs minimizer is the unique connecting committee.
\end{theorem}
%\rmr{It looks like we could use another simple measure here of $V(C,E)$ which is the number of voters in the connected component (so it is some rough proxy of the global measures). Then the theorem can be something like `for any $\delta?0$ there is E s.t. for (unique) $\hat C= PAIRS(E)$ and $\check C=\argmin_C PAIRS(C,E)$, we have $|V(\hat C,E)|<\delta n$, while $|V(\check C,E)|=n$. Then explain intuitively that the locally optimal committee is the worst globally and vice versa.  }
%\rmr{figure?}
The construction contrasts two size-$\dsmodulus$ committees $X$ and $Y$.  We choose $\dsmodulus=\dsk(\dsk-1)+1$ from a Singer perfect difference set~\cite{singer1938theorem} and split the electorate into a small population of size $\varepsilon n$ and a large population of size $(1-\varepsilon)n$, where $\varepsilon<\delta$.  The $X$-candidates create a highly regular pattern of dense overlap inside the small population: the perfect-difference-set property ensures that every pair of small voter groups is linked by exactly one $X$-candidate.  The $Y$-candidates provide the bridges that connect the large population and attach the small voters, making $Y$ the unique connecting committee.

The parameters are chosen so that these global bridges are expensive in the \pairs: replacing any $X$- by $Y$-member strictly decreases the number of directly linked voter pairs. Hence $X$ is the unique \pairs-optimal committee.  Yet under $X$ only the $\varepsilon n$ voters in the small population form a nontrivial connected component, while every voter in the large population is isolated.  Since $\varepsilon$ can be arbitrarily small, the largest connected component of the \pairs optimum can contain an arbitrarily small fraction of the electorate.  The full construction and edge-counting proof are given in Appendix~\ref{app:pairs-vanishing-component}.

Dong et al.'s Example~3.1 shows that \pairs can fail to distinguish a connected committee from a disconnected one. Theorem~\ref{thm:pairs_vanishing_component} gives a stronger asymptotic failure: even when a committee connecting all voters exists, the unique \pairs optimum may interlace only an arbitrarily small fraction of the electorate, favoring dense local connectivity over bridges necessary for global interlacing.

\subsection{CONS suboptimal for bottlenecks}
In contrast to \pairs, \cons is, as already observed by \citeauthor{dong2025selecting}, a global connectivity objective. Indeed, every connecting committee attains the maximum possible \cons score. The following observation shows that \cheeger shares this basic connectivity sensitivity.

\begin{observation}
    \label{obs:disconnected_cheeger0}
    Let $G$ be an undirected graph. Then $\Cheeger{G}=0$ iff $G$ is disconnected.
\end{observation}
\begin{corollary}
    \label{cor:cheeger_connected_committees}
    For a fixed election $\Election$:
    \begin{enumerate}
        \item the $\cheeger$ score does not distinguish between committees whose committee graphs are disconnected; and
        \item if a connecting committee exists, every $\cheeger$-optimal committee is connecting.
    \end{enumerate}
\end{corollary}

Thus, both \cons and \cheeger return a connecting committee whenever one exists. However, since all connecting committees achieve the same maximal \cons score, it cannot distinguish between them. On the other hand, \cheeger exhibits greater granularity among connecting committees, as demonstrated by this example.

\begin{example}
\label{ex:cons-cheeger-bottleneck}
Let $n=2x$ with $x\ge2$, let the candidate set be $\braces{a,b,c,d}$, and let $\CommitteeSize=3$. The voters are \rmr{transpose the table to save space}\gbs{V}
\begin{center}
\begin{tabular}{c|c|c|c|c}
approval set
& $\braces{a,d}$
& $\braces{a,c,d}$
& $\braces{b,c,d}$
& $\braces{b,d}$\\
\hline
number of voters
& $x-1$
& $1$
& $1$
& $x-1$
\end{tabular}
\end{center}
For $\Committee=\braces{a,b,c}$, the committee graph consists of two copies of $K_x$ joined by a single bridge edge. Every committee containing $d$ induces $K_{2x}$, since every voter approves $d$. Hence all four committees are connected and have the same maximal \cons score $\binom{2x}{2}$.

For the polarizing committee, all of $K_x$ gives a cut with one crossing edge. Connectedness gives the matching lower bound, so $\cheeger{\braces{a,b,c},\Election}=\frac1x$.
For a complete graph $K_{2x}$, a side of size $s\le x$ has expansion $2x-s$, minimized at $s=x$, and hence $\Cheeger{K_{2x}}=x$.
Thus \cheeger strictly prefers each committee containing $d$ to $\braces{a,b,c}$, while \cons ties all four. In particular, under some tie-breaking \cons may select the committee whose graph has a single fragile bridge between two dense voter groups.

Moreover, the polarized committee attains the minimum possible positive \cheeger score for any $n$ voters, since $\forall S\ne\emptyset,\Boundary{S}\ge 1$ and $x$ is the largest possible smaller side subset. Also, $K_n$ attains maximal expansion for every graph over $n$ vertices. We therefore obtain a tight $\Theta\left(n^2\right)$ worst case ratio between the best and worst \cheeger scores of a given election.

\begin{figure}[t]
    \centering

    \begin{subfigure}[t]{0.49\columnwidth}
        \centering
        \begin{tikzpicture}[
            scale=0.60,
            every node/.style={circle,draw,fill=white,inner sep=1.55pt},
            graphEdge/.style={line width=0.40pt},
            bridgeEdge/.style={line width=1.25pt},
            cutLine/.style={densely dashed,line width=0.75pt},
            candidateMark/.style={gray,dashed,line width=0.7pt,rounded corners=7pt}
        ]
            % Two schematic copies of K_x.  Visible vertices are only
            % representatives; the ellipses indicate omitted vertices.
            \coordinate (LC) at (1.35,1.65);
            \coordinate (RC) at (4.65,1.65);

            % Left clique: six representative vertices, with a gap below.
            \node (L1) at ($(LC)+(0:0.92)$) {};
            \node (L2) at ($(LC)+(51.43:0.92)$) {};
            \node (L3) at ($(LC)+(102.86:0.92)$) {};
            \node (L4) at ($(LC)+(154.29:0.92)$) {};
            \node (L5) at ($(LC)+(205.71:0.92)$) {};
            \node (L6) at ($(LC)+(257.14:0.92)$) {};

            % Right clique: exact mirror image of the left clique.
            \node (R1) at ($(RC)+(180:0.92)$) {};
            \node (R2) at ($(RC)+(128.57:0.92)$) {};
            \node (R3) at ($(RC)+(77.14:0.92)$) {};
            \node (R4) at ($(RC)+(25.71:0.92)$) {};
            \node (R5) at ($(RC)+(334.29:0.92)$) {};
            \node (R6) at ($(RC)+(282.86:0.92)$) {};

            % Complete all visible pairs inside each schematic clique.
            \foreach \i in {1,...,6}{
                \foreach \j in {1,...,6}{
                    \ifnum\i<\j
                        \draw[graphEdge] (L\i)--(L\j);
                        \draw[graphEdge] (R\i)--(R\j);
                    \fi
                }
            }

            % Three dots continue the circular arrangement and indicate
            % omitted vertices.  The two dot patterns are exact reflections.
            \foreach \ang in {291.43,308.57,325.71}
                \fill ($(LC)+(\ang:0.92)$) circle (0.75pt);
            \foreach \ang in {214.29,231.43,248.57}
                \fill ($(RC)+(\ang:0.92)$) circle (0.75pt);

            % Unique bridge and the corresponding sparse cut.
            \draw[bridgeEdge] (L1)--(R1);
            \draw[cutLine] (3.00,0.30)--(3.00,3.00);

            % Candidate c is approved exactly by the two bridge voters.
            % The dashed enclosure marks the candidate-induced connection
            % without competing visually with the minimum-expansion cut.
            \draw[candidateMark] (3.00,1.65) ellipse [x radius=0.94, y radius=0.31];
            \node[draw=none,fill=white,inner sep=0.8pt,text=gray] at (3.63,2.03) {$c$};

            \node[draw=none,fill=none] at (1.35,0.22) {$K_x$};
            \node[draw=none,fill=none] at (4.65,0.22) {$K_x$};
        \end{tikzpicture}
        \caption{Committee $\{a,b,c\}$ --- two dense communities connected by a weak bridge; $|\partial S|=1$ and \textsc{Cheeger} value $1/x$.}
        \label{fig:cons-cheeger-bridge}
    \end{subfigure}
    \hfill
    \begin{subfigure}[t]{0.49\columnwidth}
        \centering
        \begin{tikzpicture}[
            scale=0.75,
            every node/.style={circle,draw,fill=white,inner sep=1.55pt},
            graphEdge/.style={line width=0.36pt},
            cutLine/.style={densely dashed,line width=0.75pt}
        ]
            % Schematic complete graph K_{2x}.  Ten visible vertices are
            % shown, with an ellipsis indicating arbitrarily many more.
            \coordinate (C) at (2.75,1.65);
            \node (V1) at ($(C)+(0:1.35)$) {};
            \node (V2) at ($(C)+(36:1.35)$) {};
            \node (V3) at ($(C)+(72:1.35)$) {};
            \node (V4) at ($(C)+(108:1.35)$) {};
            \node (V5) at ($(C)+(144:1.35)$) {};
            \node (V6) at ($(C)+(180:1.35)$) {};
            \node (V7) at ($(C)+(216:1.35)$) {};
            \node (V8) at ($(C)+(252:1.35)$) {};
            \node (V9) at ($(C)+(288:1.35)$) {};
            \node (V10) at ($(C)+(324:1.35)$) {};

            % Complete all visible pairs.
            \foreach \i in {1,...,10}{
                \foreach \j in {1,...,10}{
                    \ifnum\i<\j
                        \draw[graphEdge] (V\i)--(V\j);
                    \fi
                }
            }

            % Gap in the schematic vertex set: K_{2x} has arbitrarily many
            % vertices, not only the representatives drawn here.
            \node[draw=none,fill=white,inner sep=1.0pt] at ($(C)+(270:1.35)$) {$\cdots$};

            % A balanced minimum-expansion cut.
            \draw[cutLine] (2.75,0.05)--(2.75,3.25);
            \node[draw=none,fill=none] at (2.75,-0.03) {$K_{2x}$};
        \end{tikzpicture}
        \caption{Any committee containing $d$ --- all voters form one cohesive collective; $|\partial S|=x^2$ and \textsc{Cheeger} value $x$.}
        \label{fig:cons-cheeger-complete}
    \end{subfigure}

    \caption{Two committee graphs induced by the same preferences graph. Both committees attain the maximal \textsc{CONS} value $\binom{2x}{2}$, but their \textsc{Cheeger} values differ by a quadratic factor. The dashed lines indicate minimum-expansion cuts.\rmr{can put graphs side by side for space}\rmr{I meant in a single column!}%\vspace{-4mm}
    }
    \label{fig:cons-vs-cheeger}
\end{figure}
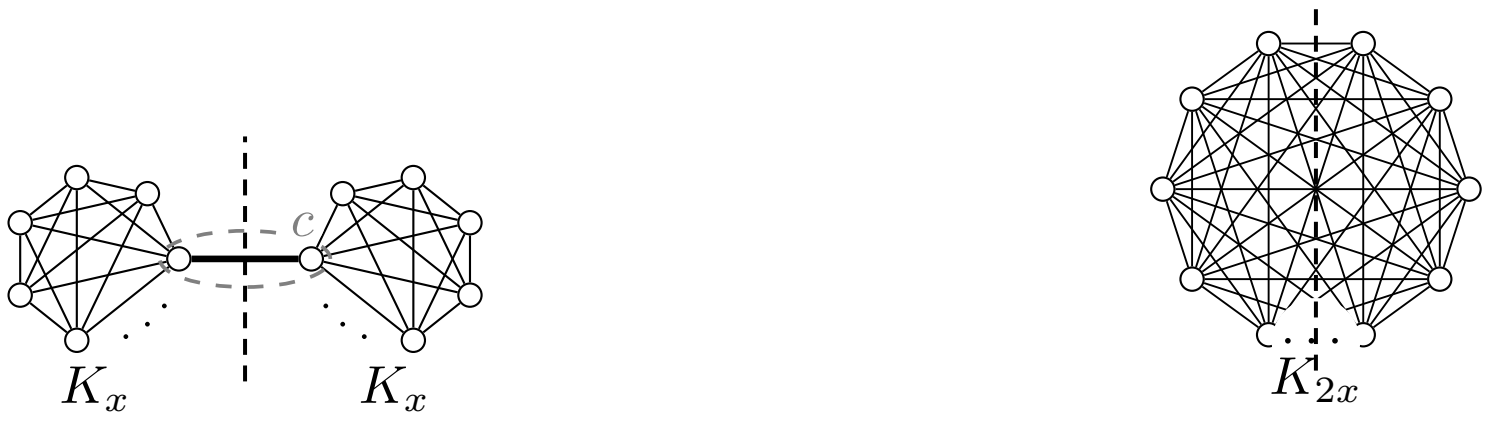

\end{example}

\subsection{CONS vs. CHEEGER on \CIDomain{} profiles}
\label{subsec:cons-cheeger-ci-separation}

On unrestricted profiles, Example~\ref{ex:cons-cheeger-bottleneck} shows that
\cons may tie committees whose \cheeger scores differ quadratically in the
number of voters.  On \CIDomain{} profiles the geometry rules out such a gap, but a linear gap remains possible.  We first show the upper bound (proof in Appendix~\ref{app:cons-subopt-proofs}).

\begin{lemma}[\CIDomain{} gateway lemma]
\label{lem:ci-gateway}
Let $\Election$ be a \CIDomain{} election on $n\ge2$ voters. Fix a connecting committee $\Committee$ and nonempty $S\subseteq\VoterSet$ with
$\abs{S}\le n/2$.  Put $B=\abs{\Boundary[\CommitteeGraph{\Committee}]{S}}$.
Let $R_S\subseteq S$ and $R_{\bar S}\subseteq\VoterSet\setminus S$ be the
vertices incident to these $B$ boundary edges.  Then every edge crossing the cut
$S\mid\bar S$ in %the preferences graph 
$\PreferencesGraph[\Election]$ has at least one endpoint in $R_S\cup R_{\bar S}$.
\end{lemma}

%\begin{proof}
%Because the selected committee is connecting, consecutive selected candidates in the CI order have overlapping supports. We consider an edge created by the joint approval of some candidate. For selected candidates consequence is immediate from definition. Any unselected candidate lies between two consecutive selected candidates\rmr{or what?}\gbs{removed}, and CI forces its support into the union of their supports; therefore, any edge it creates across the cut must touch a vertex already incident to a boundary edge of the selected committee. 
% Complete details are given in Appendix~\ref{app:cons-subopt-proofs}.
% \end{proof}

\begin{theorem}
\label{thm:ci-connected-n-approx}
Let $\Election$ be a \CIDomain{} election on $n\ge2$ voters.  For every connecting
committee $\Committee$ and every feasible committee $\CommitteePrime$,
\[
    \cheeger{\CommitteePrime,\Election}
    \le
    n\,\cheeger{\Committee,\Election}.
\]
Consequently every connecting committee is an $n$-approximation for the
\cheeger objective.
\end{theorem}

\begin{proof}
Let $S$ be a smaller side of a \cheeger-optimal cut of
$\CommitteeGraph{\Committee}$, and write $s=\abs{S}, B=\abs{\Boundary[\CommitteeGraph{\Committee}]{S}}$.
Thus $1\le s\le n/2$ and
\(
    \cheeger{\Committee,\Election}=\frac{B}{s}
\). 
Let $R_S$ and $R_{\bar S}$ be as in Lemma~\ref{lem:ci-gateway}.  Since each
boundary edge has exactly one endpoint on each side, $\abs{R_S},\abs{R_{\bar S}}\le B$.
By Lemma~\ref{lem:ci-gateway}, every edge of
$\Boundary[\CommitteeGraph{\CommitteePrime}]{S}$ has an endpoint in
$R_S\cup R_{\bar S}$.  Hence
\[
\begin{aligned}
    \abs{\Boundary[\CommitteeGraph{\CommitteePrime}]{S}}
    \le
    \abs{R_S}(n-s)+\abs{R_{\bar S}}s
    \le B(n-s)+Bs
    =Bn.
\end{aligned}
\]
Evaluating $\CommitteePrime$ on this particular cut gives $\cheeger{\CommitteePrime,\Election}
    \le
    \frac{Bn}{s}
    =
    n\,\cheeger{\Committee,\Election}$.\qedhere
\end{proof}

Theorem~\ref{thm:ci-connected-n-approx} applies in particular to every
\cons-optimal committee whenever a connecting committee exists.  Combined with the standard dominance preprocessing of \VCIDomain{}
profiles, it also applies to a connecting committee selected from the reduced \CIDomain{} instance.

The linear factor is asymptotically tight.  Moreover, it remains tight even
under the natural secondary criterion of maximizing \pairs among \cons-optimal
committees.

\begin{theorem}[\cons vs \cheeger lower bound]
\label{thm:cons-pairs-linear-cheeger-gap}
For infinitely many $n$, there exists a dominance-free \CIDomain{} election with two \cons-optimal committees $\Committee_B, \Committee_G$ such that
\(
    \frac{\cheeger{\Committee_G,\Election}}
         {\cheeger{\Committee_B,\Election}}
    =\Theta(n).
\)
\end{theorem}
\begin{proof}
Let $\Election$ be an election with $k=4$, candidates ordered $a<b<c<d<e$,
and let the voter multiplicities be \rmr{transpose or split to two columns}\gbs{v}
\begin{center}
\begin{tabular}{c|c||c|c}
number of voters & approval set & number of voters & approval set\\
\hline
$1$     & $\braces{a}$      &   $t$     & $\braces{c,d}$\\
$t+1$   & $\braces{a,b}$    &   $1$     & $\braces{c,d,e}$\\
$t$     & $\braces{a,b,c}$  &   $t$     & $\braces{d,e}$\\
$t+1$   & $\braces{b,c}$    &   $1$     & $\braces{e}$
\end{tabular}
\end{center}
Every ballot is an interval of the displayed candidate order, so the profile
is \CIDomain{}.  The five candidate supports are pairwise incomparable, so no candidate
is dominated. Set $\Committee_B=\braces{a,b,c,e},
    \qquad
    \Committee_G=\braces{a,c,d,e}$.
These are the only connecting size-$4$ committees.  Omitting $a$ isolates the
$\braces{a}$ voter, omitting $e$ isolates the $\braces{e}$ voter, and omitting
$c$ separates the voters represented by $a,b$ from those represented by
$d,e$.  By contrast, both $\Committee_B$ and $\Committee_G$ connect all
voters.  Thus both are \cons-optimal.

Next, under $\Committee_B$, let $S_B$ consist of the $t$
voters of type $\braces{d,e}$ together with the single voter of type
$\braces{e}$.  In $\CommitteeGraph{\Committee_B}$ these $t+1$ voters are
joined to the rest of the graph only through the single voter of type
$\braces{c,d,e}$, and each has exactly one edge across this cut.  Therefore $\cheeger{\Committee_B,\Election}
    \le
    \Expansion[\CommitteeGraph{\Committee_B}]{S_B}
    =1$.

For \(\Committee_G\), partition the voters into six consecutive blocks so that the committee graph contains a spanning blow-up of a six-vertex path, with complete bipartite graphs between consecutive blocks. A case analysis on how a cut intersects these blocks gives a matching linear lower bound on every cut, while a singleton cut gives the corresponding upper bound; hence 
\(
\left( \cheeger{\Committee_G,\Election}
=\Theta(t) \right)=\Theta(n)
\).
Details appear in Appendix~\ref{app:cons-subopt-proofs}.

From the linear bound and
$\cheeger{\Committee_B,\Election}\le1$, we get
\[
    \frac{\cheeger{\Committee_G,\Election}}
         {\cheeger{\Committee_B,\Election}}
    =\Omega(n).
\]
The matching $O(n)$ bound follows from
Theorem~\ref{thm:ci-connected-n-approx}.%, completing the proof.
\end{proof}

Theorem~\ref{thm:cons-pairs-linear-cheeger-gap} suggests a worst-case asymptotic order for general \cons. The following corollary demonstrates that this is not solely the result of a carefully crafted adversarial selection but rather the result of a natural \cons tie-breaking scheme using \pairs.

\begin{corollary}
\label{cor:cons-cheeger-tight-ci}
On \CIDomain{} profiles that admit a connecting committee, the worst-case multiplicative separation between the \cheeger score of a \cons-optimal with \pairs tie-break committee
and the \cheeger optimum is $\Theta(n)$.
\end{corollary}
\begin{proof}
    Let $\Committee_B, \Committee_G$ as in the proof of Theorem~\ref{thm:cons-pairs-linear-cheeger-gap}. We show that $\pairs{\Committee_B,\Election}
        >
        \pairs{\Committee_G,\Election}$.
    Start from the common subcommittee $\braces{a,c,e}$. Adding $b$ creates exactly the edges between the $t+1$ voters of type $\braces{a,b}$ and the $t+1$ voters of type $\braces{b,c}$ that were not already covered by $a$ or $c$. The marginal \pairs contribution is therefore $(t+1)^2$.
    Adding $d$ instead creates exactly the edges between the $t$ voters of type
    $\braces{c,d}$ and the $t$ voters of type $\braces{d,e}$ that were not already
    covered by $c$ or $e$, for a marginal contribution $t^2$.
    Consequently, $\pairs{\Committee_B,\Election}
        -
        \pairs{\Committee_G,\Election}
        =2t+1>0$.
    Thus, \pairs tie-breaking selects $\Committee_B$.
\end{proof}

\newpar{Interpretation.}
The result separates two notions that coincide at the level of mere
connectivity.  \cons assigns every connecting committee the same maximum score,
whereas \cheeger distinguishes how robustly the electorate is connected.  The
\CIDomain{} structure limits the damage: every connecting committee is automatically an
$n$-approximation to \cheeger.  Nevertheless this guarantee is tight, and even
a natural secondary objective that rewards additional direct voter links---
\pairs---can force the asymptotically worst-order \cheeger outcome.  Thus the
linear gap is not an artifact of adversarial tie-breaking; it persists under a
canonical refinement using the other interlacing objective.

%%%%%%%%%%%%%%%%%%%%%%%%%%%%%%%%%%%%%%%%%%%%%%%%%%%%%%%%%%%%%%%%%%%%%%%%

\section{Information Diffusion}
\label{sec:information-diffusion}
\rmr{what can you formally write as a proposition? also separate to known background and what you add.   define formally in the election notation what is the desired property (time to consensus? under what dynamics?) then prove this bounded from both sides by cheeger via translation of the terms to graph theory terms and using known results.}\gbs{Massive rework for this section}\rmr{much better. can be shorter, see comments }
In this section we formalize a secondary interpretation for the interlacing model using the DeGroot model
of opinion aggregation~\cite{degroot1974reaching,berger1981necessary}
(see also~\cite{jackson2008degroot} for a broader overview).
In a DeGroot process, every agent holds a real-valued opinion. At each
synchronous round, agents update their opinions to a weighted average of opinions of the individuals with whom they interact. Interactions are described by a fixed underlying network, with edges weighted by the relative influence individuals have on each other.

We interpret the preferences graph as an approximation of the underlying social interaction network among voters. This interpretation is supported by empirical evidence of a correlation between political similarity and social relationships, including interaction frequency~\cite{bond2015quantifying,algan2026friendship}. Indeed, the correspondence between political orientation and network community structure was demonstrated in early studies~\cite{adamic2005political,newman2006modularity}, and community detection has since become an established tool for studying political polarization~\cite{garimella2018quantifying,interian2023network,zhang2026measuring}.

We first review the relevant Markov-chain and DeGroot background. We then define a DeGroot process induced by a selected committee and characterize when it reaches global consensus. Finally, we relate the convergence rate of this process to the \cheeger objective through conductance and mixing time.

\rmr{I have a brilliant idea for your next paper :)  the voters are points in [0,1] and we can think of different committees as `moving' these points (e.g. via on-step Interlacing DeGroot or a similar model). Then the mean/max distance of points at the end of the process is a measure of polarization. }\gbs{tl;dr XD I'll observe more carefully after the submission:)}
\gbs{Appendix candidate}\rmr{ok but keep the definitions you use later}\gbs{bye bye}

\todo{this transition needs to be smoothed out.}
Extended formal background can be found in Appendix~\ref{app:diffusion-background}.

\subsection{Interlacing DeGroot Model}

We model interactions through the committee graph, so that only connections supported by the selected committee contribute to the diffusion process.
For an election $\Election$ with $n\ge 2$ voters, and a committee $\Committee$, we describe a symmetric and homogeneous influence diffusion process where `similar' voters (who approve similar candidate) repeatedly meet and average their positions. Let $\Laplacian{\Committee} = D-A$
be the combinatorial Laplacian of
$\CommitteeGraph[\Election]{C}$, with $D$ its diagonal degree matrix and $A$ its adjacency matrix. For a global susceptibility parameter
$\beta\in(0,1)$, we consider
\[
    \DiffusionMatrix[\beta]{\Committee}
    :=
    I-\frac{\beta}{n-1}\Laplacian{\Committee}.
\]
Equivalently, every edge incident to voter $i$ receives influence weight
$\beta/(n-1)$, while voter $i$ retains self-weight
\(
    1-\frac{\beta d_i}{n-1},
\)
where $d_i$ is the degree of $i$ in $G_{\Committee}$. 
\rmr{this detail can go to appendix:}
The normalization by $n-1$ uses the largest possible degree of a
committee-induced graph and therefore applies uniformly to all committees. Constant-edge-weight iterations of the form $I-\alpha L$ are commonly used in distributed averaging and consensus~\cite{olfati2004consensus,xiao2004fast}.
\rmr{up to here it's just the standard process of creating the matrix from the graph, right? So this can be part of the background with $P^\beta_H$ where $H$ is some graph (no committees involved, and you can treat it as some standard black box). then here you define $H=G^E_C$ and state the theorems and proofs with $P^\beta_H$ }\gbs{Not exactly because there is no standard selection of weights in DeGroot model and definitely not for Markov processes so really the point is the weighting and it's the only real modeling part other than "let's DeGroot it"}

For the mixing-time analysis we take $\beta=1/2$ and abbreviate
$\DiffusionMatrix{\Committee}:=\DiffusionMatrix[1/2]{\Committee}$. If
\(
    Q_{\Committee}
    =
    I-\frac{1}{n-1}\Laplacian{\Committee},
\)
then
\[
    \DiffusionMatrix[1/2]{\Committee}
    =
    \frac12\left(I+Q_{\Committee}\right),
\]
the standard half-lazification of $Q_{\Committee}$.
Lazification maps every eigenvalue $\lambda\in[-1,1]$ to
$(1+\lambda)/2\in[0,1]$ and guarantees self-weight at least $1/2$~\cite{levin2026markov}.

As a concrete DeGroot model, our results mostly are a consequences of standard DeGroot and Markov process analysis and are therefore omitted. For completeness, they are included in the supplementary material.\gbs{??????}\rmr{which appendix?}  \rmr{$t_{mix}$ not defined! }

\begin{theorem}
\label{thm:diffusion-disconnected-divergence}
    For $\DiffusionMatrix[\beta]{\Committee}$, each connected component converges to the arithmetic mean of initial opinions in that component. If $\Committee$ is connecting then convergence is to the arithmetic mean of all voters.
\end{theorem}

Consequently, if $G_{\Committee}$ is disconnected, different components may converge to distinct local consensus values.
\begin{corollary}
    A \pairs-maximizing committee may lead to a fragmented population with distinct local consensus values, even when another feasible committee would lead to global consensus.
\end{corollary}
\newpar{Proof idea.} Take an election in which the \pairs-maximizer is disconnected although a connecting committee exists. Assign a distinct initial opinion, constant within each connected component of the \pairs-maximizer. That opinion profile is already a non-consensus steady state for the \pairs-maximizer, whereas the connecting committee converges to global consensus regardless of the initial opinion distribution.

\newpar{Bounded time to consensus.}
We now obtain a direct connection between the time to reach consensus, equivalent ti the mixing time $t_{mix}$ of a Markov chain, and the \cheeger objective. We assume for the rest of this analysis that $\Committee$ is connecting.
\rmr{this is the key theorem of the section, but not clear what $t_{mix}$ stands for exactly in the context of voting. You defined it  for the deGroot process but that's part of the background and not the connection to voting.
If we needed to cut the section to a single paragraph, it would be something like: `here are voters on a (political) spectrum. suppose they communicate via such and such process (each time a pair that voted for a shared candidate average?). For a committee C we define $t_{agree}(C)$ as the time it takes to reach almost-consensus...'  then theorem 4.3 and a proof sketch saying that the process we defined is just the deGroot dynamics on $G_C$ (or $P_C$ or whatever), and that time to consensus equals mixing time of the graph, which is bounded by conductance, hence by Cheeger.    }

\begin{theorem}
\label{thm:diffusion-bounded-consensus}
    For the lazy diffusion process, the \cheeger score controls the time to consensus through its bounds on mixing time:
    \[
        t_{mix} = \Omega\parentheses{\frac{n}{\Cheeger{\CommitteeGraph{\Committee}}}},\qquad t_{mix}=O\parentheses{\frac{n^2\log n}{\Cheeger{\CommitteeGraph{\Committee}}^2}}
    \]
\end{theorem}
Importantly, the dependence on $n$ is common to all committees in a fixed election and is therefore not committee-specific.

\rmr{too much space for this example, just say what would be the mixing time (as a function of n or just for some fixed n) for each one of the graphs you already show}
Example~\ref{ex:cons-cheeger-bottleneck} illustrates the gap between consensus time of different connecting committees. 
For $n=2x$, there are two committees considered. The first induces a complete graph of all $2x$ voters, and the second induces two copies of $K_x$ joined by one bridge. Both committees are connected and indistinguishable by \cons.
$K_{2x}$ is the \cheeger-maximizer with value $x$, and the polarized bridge graph has \cheeger value of $1/x$. The bounds let us assert that the time to consensus ratio is at least \rmr{hard to follow. why not just add the mixing time of each committee for one or two fixed $n$?}\gbs{because you get a non tight bound. I rephrased a bit, hope that's better. But I made an APP}
\(
    \Omega\parentheses{\frac{\frac{n}{1/x}}{\frac{n^2\log n}{x^2}}} = \Omega\parentheses{\frac{x^3}{n\log n}} = \Omega\parentheses{\frac{n^2}{\log n}}.
\)

Thus, in settings where time to reach consensus is limited or otherwise significant, the choice of a connecting committee can substantially affect how quickly consensus is approached, with \cheeger serving as a quantitative proxy for convergence time. An interactive page demonstration of the convergence dynamics is included in the supplementary material.

%%%%%%%%%%%%%%%%%%%%%%%%%%%%%%%%%%%%%%%%%%%%%%%%%%%%%%%%%%%%%%%%%%%%%%%%

\section{Complexity}
\label{sec:complexity}

We now turn to the computational cost of bottleneck-sensitive interlacing.
The picture is negative at several levels: even obtaining any positive
\cheeger score is hard, evaluating the score of one fixed committee is hard, and committee selection remains hard under voter-interval structure.  We then show that bounded committee size yields an exact XP algorithm in general, while \VCIDomain structure substantially improves its dependence on the committee size.\rmr{I don't see the latter result in the paper?}\gbs{it wasn't in a theorem environment, it is now.}

\subsection{General Hardness}
\label{subsec:cheeger-general-hardness}

We begin with the weakest nontrivial selection question.  By
Observation~\ref{obs:disconnected_cheeger0}, a committee has positive \cheeger score exactly when its committee graph is connected.

\begin{theorem}
\label{cor:positive_cheeger_committee_npcomplete}
It is NP-complete to decide whether an election admits a size-$\CommitteeSize$
committee $\Committee$ with $\cheeger{\Committee,\Election}>0$.
Consequently, outputting a \cheeger-optimal committee is NP-hard.
\end{theorem}

\begin{proof}
Dong et al.~\cite{dong2025selecting} prove NP-completeness of deciding
whether a size-$\CommitteeSize$ committee connects all voters.  By
Observation~\ref{obs:disconnected_cheeger0}, this is equivalent to asking
whether some size-$\CommitteeSize$ committee has positive \cheeger score.
Membership is immediate by guessing a committee and checking connectivity. Moreover, on the same instance, the optimal \(\cheeger\) value is positive iff a connecting size-\(\CommitteeSize\) committee exists. Hence, a polynomial-time algorithm for  a \(\cheeger\)-optimal committee would decide the NP-complete problem.
\end{proof}

\citeauthor{dong2025selecting} show that generally interlacing is NP-hard for both \pairs and \cons. However, while these objectives admit polynomial-time calculation of a fixed committee's value and hardness is the result of the exponential enumeration of possible committees, even evaluating one committee is computationally difficult. Recall that the decision variant of computing the \cheeger value of a general graph is known to be NP-complete. The following shows that this extends to committee-induced subgraphs as well.

\begin{theorem}
\label{thm:fixed_committee_cheeger_npcomplete}
For an election
$\Election=\parentheses{\CandidateSet,\VoterSet,\ApprovalProfile,\CommitteeSize}$,
a committee $\Committee$ of size $\CommitteeSize$, and a rational threshold
$\Threshold>0$, deciding whether $\cheeger{\Committee,\Election}\le\Threshold$
is NP-complete.
\end{theorem}

The reduction is direct.  Given a graph $G=(V,E)$, create one voter for each
vertex and one candidate for each edge, approved by exactly its two endpoint
voters, and select the full candidate set.  The resulting committee graph is
exactly $G$, so ordinary graph-Cheeger decision is reproduced without
distortion.  The full formal proof is given in
Appendix~\ref{app:fixed-committee-hardness}.

\subsection{Hardness on Voter-Interval Profiles}
\label{subsec:vi-cheeger-hardness}

\citeauthor{dong2025selecting} manages to mitigate the general hardness result by analyzing restricted preference domains, and obtaining polynomial-time algorithms for both objectives in the \VCIDomain{} domain. Unfortunately, such a restriction does not help for \cheeger.

\begin{theorem}[\VIDomain{} hardness]
\label{thm:vi_cheeger_exact_k_nphard}
Given a voter-interval election $\Election$, an integer
$\CommitteeSize$, and a rational threshold $\Threshold>0$, it is NP-hard to decide
whether there exists a committee $\Committee$ with
$\abs{\Committee}=\CommitteeSize$ and $\cheeger{\Committee,\Election}\ge\Threshold$.
\rmr{this detail can go in appendix:}The hardness holds even when all candidate supports are nonempty and pairwise
distinct, and the support family is proper (no support properly contains another).
\end{theorem}

\gbs{If/How to cite the supplementary material?}

\newpar{Reduction overview.}
We reduce from 
\StrictUnaryThreePartition, in which a set of weighted items is asked to be partitioned into 3-item bins of a predefined target weight $\VIHSourceTarget$. The problem has the additional constraint that each item's weight is bounded $\frac{\VIHSourceTarget}{4}
    <
    \VIHSourceWeight{i}
    <
    \frac{\VIHSourceTarget}{2}$, thus ensuring that a set of items with the target weight is composed of exactly 3 items. \StrictUnaryThreePartition is known to be NP-complete, even when input is unary-represented~\cite{garey2002computers}.

The construction encodes a solution as a sequence of cumulative sets of items referred to as \emph{stages}. A stage $h$ encodes all items assigned to the first $h$ bins of the solution. For each $h<\VIHSourceBins$, stage $h$ is represented by two \emph{variable phases}, $2h$ and $2h+1$, which respectively impose a lower and an upper bound on the total weight of the items in the stage.

For each phase $p$ and item $i$, there are two alternative candidates $\VIHBranchA{p}{i}$ and $\VIHBranchB{p}{i}$. The construction will force exactly one of them to be selected. We write
\(
\VIHChoice{p}{i}=
\begin{cases}
1,&\VIHBranchA{p}{i}\text{ is selected},\\
0,&\VIHBranchB{p}{i}\text{ is selected},
\end{cases}
\)

The interpretation is that $\VIHChoice{p}{i}=1$ once item $i$ has entered the cumulative collection of completed 3-PARTITION bins. This monotone property is enforced by phase- and item-specific markers, specific prefix cuts. Thus any committee meeting the threshold ensures $\VIHChoice{p}{i}=1 \rightarrow \VIHChoice{p'}{i}=1$ for every subsequent phase $p'$.

Additionally, for each phase $p$ and item $i$, two auxiliary weighting candidates, an anchor $\VIHItemAnchor{p}{i}$ and a weight-dependent component $\VIHMergeCandidate{p}{i}$ are introduced. Together with the two alternative candidates, they create an $a_i$-sized multiplier for $\VIHChoice{p}{i}$. This is achieved using a weight probe, a single phase-specific selected prefix, used to measure the cumulative weight effect of all items. At the weight probe, each individual item's weighting candidates create an $a_i$ difference in boundary by choosing $\VIHBranchA{p}{i}$ relative to choosing $\VIHBranchB{p}{i}$. Consequently, $\VIHEncodedWeight{p}=\sum_i \VIHSourceWeight{i} \VIHChoice{p}{i}$ is the cumulative weight encoded in phase $p$.

The two phases associated with a stage encode this same cumulative weight with opposite signs: in an even phase, selecting $\VIHBranchA{p}{i}$ contributes $+\VIHSourceWeight{i}$ to the relevant boundary expression relative to selecting $\VIHBranchB{p}{i}$, while in an odd phase it contributes $-\VIHSourceWeight{i}$. Hence the committee-dependent contribution of $p$ is proportional to $(-1)^p\VIHEncodedWeight{p}$. 

Each phase also includes a small set of calibration candidates with the intended purpose of comparing $\VIHEncodedWeight{p}$ against the desired $h\VIHSourceTarget$ cumulative weight at stage $h$. Even and odd phases impose opposite inequalities on the cumulative weight represented by that stage. The calibration candidates cause the baseline boundary of the weight probe to correspond with the target cumulative weight. The alternating sign then makes weights below the target fail in the even phase and weights above the target fail in the odd phase, so the two phases jointly impose both a lower and an upper bound on the cumulative weight represented by the stage. Therefore, any committee that guarantees the threshold must encode cumulative weight exactly $h\VIHSourceTarget$ at stage $h$. 
Consequently, the differences between stages form a valid 3-partition.

The remaining candidates have three auxiliary purposes: they calibrate all diagnostic cuts (i.e., monotonicity markers and weight probes) to one common threshold, force every feasible committee into the intended $\VIHBranchALabel/\VIHBranchBLabel$ form, and provide enough surplus to make all nondiagnostic cuts safe, i.e. have expansion above the threshold. The construction is first analyzed through prefix cuts of the voter order. After establishing the required bounds for all prefixes, we separately show that every arbitrary cut is at least as safe, using the interval structure together with large head and tail cliques.

All candidate supports are intervals whose left and right endpoints increase in the same candidate order, so they are pairwise distinct proper \VIDomain{} intervals. The complete construction and proof appear in Appendix~\ref{app:vi-hardness}. A Lean formalization certifying the correctness of the proof is included in the supplementary material.

\subsection{Tractability for Bounded Committee Size}
\label{subsec:bounded-k-cheeger}
Despite the general hardness results, we obtain tractability for bounded committee size by grouping voters according to their approval behavior with respect to the selected committee, thereby making the combinatorial dependence primarily a function of the committee size \(k\), rather than the number of voters.

Fix an election $\Election$ and a committee $\Committee$, and write
$G:=\CommitteeGraph[\Election]{\Committee}=(V,E)$.  The
\emph{committee-relative type} of a voter is $\vtype[\Committee]{v}:=\ApprovalBallot{v}\cap\Committee$.
Let $\TypeSet[\Committee]{V} = \braces{\vtype[\Committee]{v} | v\in V}$ be the set of occurring types. For every \(\alpha\in\TypeSet[\Committee]{V}\), let
$\TypeClass[\Committee]{\alpha}$ and
$\TypeWeight[\Committee]{\alpha}$ denote the corresponding voter class and
its cardinality ($\Committee$ is omitted from notation when clear from context). Voters of the same type have identical neighborhoods outside
their type class, and every nonempty type class induces a clique.

The associated weighted type graph $\TypeGraph{G}$ has $\TypeSet[\Committee]{V}$ as vertices, each type $\alpha$ has vertex weight $\TypeWeight{\alpha}$, and an edge connects distinct types
$\alpha,\beta$ exactly when $\alpha\cap\beta\ne\emptyset$, with edge weight $\TypeWeight{\alpha}\cdot\TypeWeight{\beta}$.
A cut \(S\mid\bar S\) is called \emph{type-respecting} if it does not split any type class, i.e., for every \(\alpha\in\TypeSet{V}\), either \(\TypeClass{\alpha}\subseteq S\) or \(\TypeClass{\alpha}\cap S=\emptyset\).
Thus, every type-respecting cut corresponds to a cut of the type graph with the same side size and boundary, in the usual weighted sense.

The following theorem shows that an optimal cut can always be chosen to be type-respecting, with at most one split type when the smaller side has maximum possible size.

\begin{theorem}
\label{thm:max_single_split_node}
Let $G=(V,E)$ be a committee graph and let \(c:=\left\lfloor\frac{\abs V}{2}\right\rfloor\).
There exists a Cheeger-optimal cut \(S^\star\mid\bar S^\star\) such that:
\begin{enumerate}
    \item if $\abs{S^\star}<c$, then $S^\star$ is type-respecting;
    \item if $\abs{S^\star}=c$, then $S^\star$ splits at most one
    voter type.
\end{enumerate}
\end{theorem}

\newpar{Proof idea.}
Below the half-size cap, if an optimal side splits a type, additional voters
of that type can be added with nonincreasing boundary marginal until either
the type is completed or the cap is reached; optimality forces each such
addition to preserve the optimum.  At the cap, if two types are split, the
sum of the costs of the two opposite one-for-one type swaps is nonpositive.
Since optimality makes both swap costs nonnegative, both are zero.  Repeating
a zero-cost transfer eliminates one split type at a time.  The full argument,
including the cap-swap calculation, is in
Appendix~\ref{app:type-graph-proof}.

\begin{proposition}
\label{cor:type_graph_cheeger_evaluation}
Given a weighted type graph with \(t\) committee-relative voter types, its \(\cheeger\) score can be computed exactly in \(2^{O(t)}\operatorname{poly}(t)\) time.
\end{proposition}
\begin{proof}
Let $c=\lfloor\frac{n}{2}\rfloor$. A naive algorithm enumerates all possible $2^t$ cuts. For each cut, if it has weight at most $c$, evaluate its expansion. Then, if its weight is also strictly less than $c$, iterate the remaining $t$ types, if any can be split to obtain an exact $c$-sized cut, evaluate its expansion. Discard all other cuts and return the optimal expansion found. Theorem~
\ref{thm:max_single_split_node} guarantees that at least one optimum is considered by this procedure. Each of the $2^t$ iterations can be completed in $\operatorname{poly}(t)$ time, thus obtaining the desired bound.
\end{proof}

Since a size-\(k\) committee induces at most \(2^k\) committee-relative voter types, the proposition immediately yields the following.

\begin{corollary}
    For an election with \(m\) candidates and \(n\) voters, a \(\cheeger\)-optimal size-\(\CommitteeSize\) committee can be computed in time $\binom{m}{\CommitteeSize}\,
        2^{O(2^{\CommitteeSize})}\,
        \operatorname{poly}(n,m)$. Consequently, \(\cheeger\) committee selection is in XP parameterized by \(\CommitteeSize\), and is polynomial-time solvable for every fixed \(\CommitteeSize\).
\end{corollary}

Although the above result yields polynomial-time solvability for every fixed \(k\), the dependence on \(k\) is doubly exponential and quickly becomes prohibitive. This blow-up is driven by the number of committee-relative voter types, suggesting that preference domains admitting substantially fewer types may yield much more practical algorithms. One such domain is \CIDomain{}. Thus, we obtain the following.
%\vspace{-0.15em}
\begin{theorem}
    For an election in the \VCIDomain{} domain, let \(N=n+m\) and \(\CommitteeSize=\operatorname{polylog}(N)\). A \cheeger-optimal size-$\CommitteeSize$ committee can be computed in quasi-polynomial time.
\end{theorem}
%\vspace{-0.5em}
\begin{proof}
    For any size-\(\CommitteeSize\) committee, every nonempty committee-relative type is a contiguous interval of the selected candidates. Hence there are at most $1+\binom{\CommitteeSize}{2}=O(\CommitteeSize^2)$ types, including the empty type. Consequently, \cheeger committee selection on \CIDomain{} profiles can be solved in $\binom{m}{\CommitteeSize}2^{O(\CommitteeSize^2)}
\operatorname{poly}(n,m)$
time. In particular, letting \(N=n+m\), this remains quasi-polynomial even when \(\CommitteeSize=\operatorname{polylog}(N)\). By the dominance preprocessing of \citeauthor{dong2025selecting}, the same bounds extend to \VCIDomain{} profiles.
\end{proof}

%%%%%%%%%%%%%%%%%%%%%%%%%%%%%%%%%%%%%%%%%%%%%%%%%%%%%%%%%%%%%%%%%%%%%%%%

\section{Discussion}
\rmr{limit to 3/4 page. about eps padding, felix fischer? paper on using PageRank as a voting method for tournaments}

Earlier committee-evaluation criteria and axioms are largely voter-centric: they assess a committee through properties or outcomes defined from the perspective of individual voters. Interlacing shifts the unit of analysis from individual voters to the pattern of connections among voters induced by the committee.
\pairs and \cons capture direct overlap and reachability, but our results show that neither captures the robustness of the connections bridging different parts of the electorate. %This suggests that structural robustness, rather than connectivity alone, is the relevant topological structure.
%Once voter-to-voter structure becomes the object of interest, the topological features of that structure determine which forms of interlacing we regard as desirable. PAIRS and CONS capture two natural aspects of this structure---direct overlap and indirect reachability---but our results show that neither captures how robust an already-connected electorate is against division into weakly linked communities. In particular, reachability alone does not distinguish between an electorate whose communities are connected by many reinforcing bridges and one in which they are joined only tenuously. While the original interlacing framework uses connectivity as a proxy for similarity among voters, so that greater connectivity corresponds to greater interlacing, our results suggest that the relevant topological feature is instead the structural robustness of the connections that bridge different parts of the electorate.

We use CHEEGER instead, to capture structural robustness, rather than connectivity alone, and is further justified by the information-diffusion interpretation.% model introduced in this paper provides a \emph{dual interpretation} of polarization induced by committee selection.
%In addition to the structural separation captured by the interlacing interpretation, polarization is commonly associated with antagonism or disagreement between groups and with divergence in their opinions~\cite{esteban1994measurement,dandekar2013biased}. The diffusion perspective makes the disagreement component explicit by assigning voters opinions and asking how the interaction structure induced by the selected committee affects their evolution: disconnected components may converge to distinct local opinions, while even in a connected electorate weakly connected communities may retain their disagreement for a long time before reaching consensus. CHEEGER then acquires a second interpretation within this view---it measures the severity of structural bottlenecks in the committee-induced graph.
Under the symmetric DeGroot process studied here, its proportionality to Markov-chain conductance connects this bottleneck measure to standard convergence bounds, making CHEEGER a proxy for the time scale on which the electorate approaches consensus.
%Thus, the same objective that rewards robust bridges between communities in the interlacing view also favors interaction structures in which disagreement is less persistently insulated in the diffusion view. 
%Thus, the two models provide dual perspectives on polarization induced by committee selection, tied together by the same graph objective. This duality also places CHEEGER within a broader family of sparsest-cut-like measures, such as conductance and normalized cut~\cite{levin2026markov,shi2000normalized};
%whose different normalizations may capture different aspects of the same underlying phenomenon. Among these related measures,
While there are other global measures, e.g. conductance and normalized cut~\cite{levin2026markov,shi2000normalized}, we focus on CHEEGER because its cardinality normalization evaluates the strength of cross-community connections relative to the number of voters being connected, without allowing changes in committee-induced within-community density to alter the normalization.
%other normalizations may be preferable when edge volume itself carries substantive meaning.

These appealing structural and interpretive properties come at a substantial computational cost. 
Optimization remains hard even under strong preference structure, although fixing the committee size and exploiting CI structure yield tractable regimes.
%Optimizing CHEEGER is computationally intractable in general, and this difficulty persists even under the strong structure of the VI domain. 
%Our positive results therefore do not eliminate the computational obstacle so much as identify settings in which it can be controlled
%: fixing the committee size yields polynomial-time algorithms, while CI structure permits substantially stronger compression of the problem. In CI elections, every connecting committee is within an asymptotically tight linear factor of the optimal CHEEGER value. More tellingly, such a gap can arise under a natural selection criterion: maximizing PAIRS among connecting committees can itself yield a linear-factor separation from the CHEEGER optimum. 
However, CHEEGER's close connection to sparse-cut formulations used in graph partitioning and community detection, where spectral, flow-based, and heuristic methods are commonly used in place of exact optimization~\cite{leskovec2010empirical}, provides a natural source of algorithmic ideas. Moreover, any such method that guarantees a connecting committee whenever one exists already inherits our universal linear-factor baseline, leaving room to improve over the worst-case behavior attained by an arbitrary---or even PAIRS-maximizing---connecting committee.

A second limitation is that CHEEGER's still does not distinguish between disconnected graphs. One natural way to soften this behavior is to introduce a \emph{damping factor}, assigning weight $\varepsilon\in(0,1)$ to absent edges of the committee graph while retaining weight $1$ on committee-induced edges. At $\varepsilon=0$ this recovers the original model, while for higher values
%whereas at $\varepsilon=1$ every committee induces the same weighted clique and committee choice no longer affects the graph. For intermediate values, 
$\varepsilon$ controls a substantive trade-off between \emph{within-component cohesion} and \emph{tolerance for outliers}. %Indeed, for the resulting weighted graph,
% \[
% h_\varepsilon(G)
% =
% \min_{0<|S|\le n/2}
% \left[
% (1-\varepsilon)\frac{|\partial_G(S)|}{|S|}
% +\varepsilon(n-|S|)
% \right].
% \]
% The first term captures the contribution of genuine committee-induced connections, while the regularization term is comparatively forgiving when only a small set of voters is separated but penalizes divisions involving a substantial part of the electorate more strongly. 
Thus, $\varepsilon$ may be treated as a regularization parameter. A more conservative alternative is to study the limit $\varepsilon\to0^+$, using the perturbation only to refine the hard-zero behavior while remaining as close as possible to the original CHEEGER objective. This approach has a methodological analogue in the study of PageRank limit scores as a tournament solution~\citet{brandt2007pagerank}.
%, who introduce PageRank tournament solution, where taking the damping factor toward $1$ removes the exogenous contribution and facilitates recovering a solution determined by the underlying dominance relation.

\rmr{I shortened up to here. did not touch next}
A further natural extension is to move from binary to weighted committee-induced graphs. 
Such weights have no immediate canonical interpretation under the original objectives, most notably for the reachability-based CONS objective. Our diffusion perspective makes such weighting particularly natural--
%Our perspective makes weighting considerably more natural. Structurally, different edge weights can substantially change the relevant cuts even when the underlying unweighted graph remains unchanged. More importantly, 
the diffusion framework developed here can be viewed as the homogeneous special case in which all committee-induced connections have equal strength. Heterogeneous interaction weights then provide a direct framework for defining and analyzing alternative weighting schemes through their effects on information flow and convergence. 
Weighted expansion and related cut objectives therefore provide a natural extension of interlacing from the existence and robustness of voter connections to their strength as well. More broadly, these extensions suggest viewing interlacing not as a single objective, but as a framework for studying how committee selection shapes the structure and strength of interactions across the electorate.

%%%%%%%%%%%%%%%%%%%%%%%%%%%%%%%%%%%%%%%%%%%%%%%%%%%%%%%%%%%%%%%%%%%%%%%%

%%% The next two lines define, first, the bibliography style to be 
%%% applied, and, second, the bibliography file to be used.

\bibliographystyle{ACM-Reference-Format} 
\bibliography{mybib}

%%%%%%%%%%%%%%%%%%%%%%%%%%%%%%%%%%%%%%%%%%%%%%%%%%%%%%%%%%%%%%%%%%%%%%%%

\appendix
\onecolumn
\section{Proof of the PAIRS pathology}
\label{app:pairs-vanishing-component}

For completeness, we give the full construction and proof of Theorem~\ref{thm:pairs_vanishing_component}.

\newpar{Perfect Difference Sets.}
A $\dsk$-sized subset of $\Z_{\dsmodulus}$ is said to be a \emph{$\parentheses{\dsmodulus,\dsk}$-Perfect Difference Set} if every value in $\braces{1,\ldots,\dsmodulus-1}$ has a unique expression as a difference of two elements in the set, calculated modulo $\dsmodulus$.

Example: the set $\braces{0,1,3}$ is a $\parentheses{7,3}$-perfect difference set:
\begin{align*}
    1-0&=1\\
    3-1&=2\\
    3-0&=3\\
    0-3&\equiv -3\equiv 4 \pmod 7\\
    1-3&\equiv -2\equiv 5 \pmod 7\\
    0-1&\equiv -1\equiv 6 \pmod 7
\end{align*}

For every $\dsk$ such that $\dsk-1$ is a prime power, a $\parentheses{\dsk\parentheses{\dsk-1}+1,\dsk}$-perfect difference set is known to exist~\cite{singer1938theorem}.

\begin{observation}
    From the above, it follows that we can choose $\dsmodulus$ to be arbitrarily large with an appropriate $\dsk$.
\end{observation}

\newpar{Proof of Theorem~\ref{thm:pairs_vanishing_component}.}
\paragraph{Construction.}
Choose an arbitrarily small positive rational $\varepsilon<\min\{\delta,1\}$. Select $\dsk$ such that $\dsk-1$ is a prime power and, for $\dsmodulus=\dsk\parentheses{\dsk-1}+1$,
\begin{equation}
    \label{eq:constraint1}
    \parentheses{\dsmodulus-1}\varepsilon^2-1 > 0.
\end{equation}
From the above observation, such $\dsk$ exists. Let $S$ be a $\parentheses{\dsmodulus,\dsk}$-perfect difference set.

Define two disjoint sets of candidates $X = \braces{x_0,\ldots,x_{\dsmodulus-1}}$ and $Y=\braces{y_0,\ldots,y_{\dsmodulus-1}}$, and set
\[
    \CandidateSet=X\cup Y,\qquad \CommitteeSize=\dsmodulus.
\]
Choose $n$ sufficiently large so that
\[
    \frac{n}{\dsmodulus}\in\N,\qquad
    \frac{n\varepsilon}{\dsmodulus^2}\in\N,\qquad
    \frac{(1-\varepsilon)n}{\dsmodulus}\ge2,
\]
and
\begin{equation}
    \label{eq:constraint2}
    n > \frac{\parentheses{1+\varepsilon}\dsmodulus^3}{\parentheses{\varepsilon^2\dsmodulus-1}\dsmodulus^2-\dsmodulus\parentheses{\dsmodulus-1}\varepsilon^2}.
\end{equation}
Such an $n$ exists because $\varepsilon$ is rational.
Note that the denominator is strictly positive as a consequence of Equation~\ref{eq:constraint1}:
\[
\parentheses{\varepsilon^2\dsmodulus-1}\dsmodulus^2
-\dsmodulus\parentheses{\dsmodulus-1}\varepsilon^2
> \varepsilon^2\dsmodulus^2
-\varepsilon^2\dsmodulus\parentheses{\dsmodulus-1}
=\varepsilon^2\dsmodulus>0.
\]

The voting profile is described as follows. First, we divide the population into two sets of sizes $\varepsilon n$ and $\parentheses{1-\varepsilon}n$. We then divide the large set into $\dsmodulus$ equally-sized groups of $\frac{\parentheses{1-\varepsilon}n}{\dsmodulus}$ voters, each group distinctly approving one candidate from $Y$. For every $0\le i\le \dsmodulus-1$, in the group that approves $y_i$ there is a single voter who also approves $y_{(i+1)\bmod \dsmodulus}$. This implies that every candidate in $Y$ is approved by $\frac{\parentheses{1-\varepsilon}n}{\dsmodulus}+1$ voters in this set.

Next, we divide the small set of voters into $\dsmodulus$ equally-sized groups $g_0,\ldots,g_{\dsmodulus-1}$, each with $\frac{n\varepsilon}{\dsmodulus}$ voters.
For every $0 \le i \le \dsmodulus-1$ and every $s \in S$, the group $g_{(i+s)\bmod \dsmodulus}$ approves $x_i$. Consequently, there are $\dsk$ groups that approve $x_i$.

Finally, each group is further divided into $\dsmodulus$ subgroups, each additionally and uniquely approving an candidate from $Y$. Thus, each candidate in $Y$ is approved by $\frac{n\varepsilon}{\dsmodulus^2}$ voters in each group $g_i$, and therefore the total number of voters for each candidate in the small set is $\dsmodulus\cdot\frac{n\varepsilon}{\dsmodulus^2}=\frac{n\varepsilon}{\dsmodulus}$. Hence each candidate from $Y$ is approved by a total of
\[
\frac{\parentheses{1-\varepsilon}n}{\dsmodulus}+1
+\frac{n\varepsilon}{\dsmodulus}
=\frac{n}{\dsmodulus}+1
\]
voters.

\begin{lemma}
    \label{lm:unique_X_alternative_per_group}
    Let Voter 1 from $g_i$ and Voter 2 from $g_j$, where $i\ne j$. Then they approve exactly one candidate from $X$.
\end{lemma}
\begin{proof}
    Existence.
    WLOG assume $i<j$. Since $S$ is a $\parentheses{\dsmodulus,\dsk}$-perfect difference set, there are $s_1,s_2\in S$ such that
    \[
    s_2-s_1\equiv j-i \pmod{\dsmodulus}.
    \]
    Let $t\equiv i-s_1\pmod{\dsmodulus}$ with $t\in\braces{0,\ldots,\dsmodulus-1}$. By construction, $x_t$ is approved by all groups with index $t+s$ for some $s\in S$. Since all calculations are taken modulo $\dsmodulus$,
    \begin{align*}
        t+s_1 &\equiv i-s_1+s_1\equiv i,\\
        t+s_2 &\equiv i-s_1+s_2\equiv i+j-i\equiv j,
    \end{align*}
    and therefore $g_i$ and $g_j$ both approve $x_t$.

    Uniqueness.
    Let $r$ be such that $x_r$ is approved by both $g_i$ and $g_j$. By construction, there are $s_{r_1},s_{r_2}\in S$ such that
    \[
    r+s_{r_1}\equiv i\pmod{\dsmodulus},
    \qquad
    r+s_{r_2}\equiv j\pmod{\dsmodulus}.
    \]
    We get
    \[
    j-i\equiv \parentheses{r+s_{r_2}}-\parentheses{r+s_{r_1}}
    \equiv s_{r_2}-s_{r_1}\not\equiv 0\pmod{\dsmodulus}.
    \]
    Since every nonzero residue modulo $\dsmodulus$ is uniquely expressed as the difference of two elements from $S$, we get $s_{r_1}=s_1$, $s_{r_2}=s_2$, and thus $r=t$.
\end{proof}

As there are exactly $\dsk$ groups that approve each $x_i\in X$, a consequence of Lemma~\ref{lm:unique_X_alternative_per_group} is that we can partition the set of edges between groups by the candidate in $X$. Moreover, if we consider the edges between groups to be a single edge weighted by the number of edges between each pair of groups (henceforth referred to as ``multi-edges''), then we get a partition into $\dsk$-cliques for each candidate in $X$.

Committee $Y$ connects all voters: each $y_i$ connects its large group, the designated bridge voters connect consecutive large groups around the cycle, and every small-set voter approves one member of $Y$. Moreover, $Y$ is the unique connecting committee. Indeed, if a committee omits some $y_i$, then a non-bridge voter in the corresponding large group is isolated; such a voter exists by the choice of $n$. Since $|Y|=\CommitteeSize$, every connecting committee is therefore exactly $Y$.

Since $\CandidateSet=X\cup Y$ and $\CommitteeSize=\dsmodulus$, every feasible committee is obtained from $X$ by replacing $t$ members of $X$ with $t$ members of $Y$, for some $0\le t\le\dsmodulus$. We compare all such committees.

Note that it is sufficient to show that the difference between the number of edges removed from $\CommitteeGraph{X}$ and the number of edges added to it by replacing $t$ candidates from $X$ with candidates from $Y$ is strictly positive for every $0 < t \le \dsmodulus$.

For a given $t$, let $\ell$ be the number of groups that approve none of the $X$ candidates that were not replaced.

The number of edges removed from $\CommitteeGraph{X}$ is as follows:
\begin{enumerate}
    \item Each candidate replaced from $X$ removes all edges from its respective $\dsk$-clique, since it is the only common $X$ candidate between voters in these groups. Each group has size $\frac{n\varepsilon}{\dsmodulus}$ and thus each multi-edge of the clique is weighted by $\parentheses{\frac{n\varepsilon}{\dsmodulus}}^2$. There are $\frac{\dsk\parentheses{\dsk-1}}{2}=\frac{\dsmodulus-1}{2}$ such multi-edges (the number of edges in a $\dsk$-clique).

    Total: $t\cdot\frac{\dsmodulus-1}{2}\parentheses{\frac{n\varepsilon}{\dsmodulus}}^2$.

    \item For each of the $\ell$ groups that approve none of the remaining $X$ candidates, all the edges in the group are removed. Since this is a $\frac{n\varepsilon}{\dsmodulus}$-clique, $\frac{\frac{n\varepsilon}{\dsmodulus}\parentheses{\frac{n\varepsilon}{\dsmodulus}-1}}{2}$ edges are removed per group.

    Total: $\ell\cdot \frac{\frac{n\varepsilon}{\dsmodulus}\parentheses{\frac{n\varepsilon}{\dsmodulus}-1}}{2}$.
\end{enumerate}

The number of edges added from the $t$ added $Y$ candidates is as follows:
\begin{enumerate}
    \item For each of the candidates from $Y$, there are $\frac{\parentheses{1-\varepsilon}n}{\dsmodulus}+1$ voters in the large set from the initial division. All these voters are now connected in a clique.

    Total: $t\cdot\frac{\parentheses{\frac{\parentheses{1-\varepsilon}n}{\dsmodulus}+1}\frac{\parentheses{1-\varepsilon}n}{\dsmodulus}}{2}$.

    \item For each candidate from $Y$, the $\frac{\parentheses{1-\varepsilon}n}{\dsmodulus}+1$ voters in the large set are connected to the $\frac{n\varepsilon}{\dsmodulus}$ voters in the small set.

    Total: $t\parentheses{\frac{\parentheses{1-\varepsilon}n}{\dsmodulus}+1}\frac{n\varepsilon}{\dsmodulus}$.

    \item For each multi-edge removed from $X$, each of the $t$ added candidates from $Y$ adds $\parentheses{\frac{n\varepsilon}{\dsmodulus^2}}^2$ edges: within each group, the voters are secondarily partitioned by the candidates in $Y$, and each of these sets of voters has size $\frac{n\varepsilon}{\dsmodulus^2}$.

    Total: $t^2\frac{\dsmodulus-1}{2}\parentheses{\frac{n\varepsilon}{\dsmodulus^2}}^2$.

    \item Lastly, for each of the $\ell$ groups that are unrepresented by the remaining $X$ candidates, the sets of partitioned voters who approve one of the added $Y$ candidates are connected in a clique.

    Total: $\ell\cdot t \frac{\frac{n\varepsilon}{\dsmodulus^2}\parentheses{\frac{n\varepsilon}{\dsmodulus^2}-1}}{2}$.
\end{enumerate}

We get that the difference in edges is given by
\begin{align*}
    t&\parentheses{
        \frac{\dsmodulus-1}{2}\parentheses{\frac{n\varepsilon}{\dsmodulus}}^2
        -\parentheses{
            \frac{\parentheses{\frac{\parentheses{1-\varepsilon}n}{\dsmodulus}+1}\frac{\parentheses{1-\varepsilon}n}{\dsmodulus}}{2}
            +\parentheses{\frac{\parentheses{1-\varepsilon}n}{\dsmodulus}+1}\frac{n\varepsilon}{\dsmodulus}
            +t\frac{\dsmodulus-1}{2}\parentheses{\frac{n\varepsilon}{\dsmodulus^2}}^2
        }
    }\\
    &+ \ell\parentheses{
        \frac{\frac{n\varepsilon}{\dsmodulus}\parentheses{\frac{n\varepsilon}{\dsmodulus}-1}}{2}
        -t\frac{\frac{n\varepsilon}{\dsmodulus^2}\parentheses{\frac{n\varepsilon}{\dsmodulus^2}-1}}{2}
    }.
\end{align*}

First, consider the $\ell$ term:
\begin{align*}
    &\frac{\frac{n\varepsilon}{\dsmodulus}\parentheses{\frac{n\varepsilon}{\dsmodulus}-1}}{2}
    -t\frac{\frac{n\varepsilon}{\dsmodulus^2}\parentheses{\frac{n\varepsilon}{\dsmodulus^2}-1}}{2}\\
    &\ge
    \frac{\frac{n\varepsilon}{\dsmodulus}\parentheses{\frac{n\varepsilon}{\dsmodulus}-1}}{2}
    -\dsmodulus\frac{\frac{n\varepsilon}{\dsmodulus^2}\parentheses{\frac{n\varepsilon}{\dsmodulus^2}-1}}{2}
    \qquad (t\le \dsmodulus)\\
    &=\frac{1}{2}\frac{n\varepsilon}{\dsmodulus}
    \parentheses{\frac{n\varepsilon}{\dsmodulus}-\frac{n\varepsilon}{\dsmodulus^2}}\\
    &=\frac{n^2\varepsilon^2\parentheses{\dsmodulus-1}}{2\dsmodulus^3}\ge 0.
\end{align*}

Therefore, it only remains to show that the $t$ term is strictly positive:
\begin{align*}
    \frac{\dsmodulus-1}{2}&\parentheses{\frac{n\varepsilon}{\dsmodulus}}^2
    -\parentheses{
        \frac{\parentheses{\frac{\parentheses{1-\varepsilon}n}{\dsmodulus}+1}\frac{\parentheses{1-\varepsilon}n}{\dsmodulus}}{2}
        +\parentheses{\frac{\parentheses{1-\varepsilon}n}{\dsmodulus}+1}\frac{n\varepsilon}{\dsmodulus}
        +t\frac{\dsmodulus-1}{2}\parentheses{\frac{n\varepsilon}{\dsmodulus^2}}^2
    }\\
    &= \frac{1}{2\dsmodulus^4}\parentheses{
        n^2\varepsilon^2\parentheses{\dsmodulus-1}\dsmodulus^2
        -\parentheses{\frac{\parentheses{1-\varepsilon}n}{\dsmodulus}+1}
         \parentheses{1+\varepsilon}n\dsmodulus^3
        -t\parentheses{\dsmodulus-1}n^2\varepsilon^2
    }\\
    &= \frac{n}{2\dsmodulus^4}\parentheses{
        n\parentheses{
            \parentheses{\dsmodulus\varepsilon^2-1}\dsmodulus^2
            -t\parentheses{\dsmodulus-1}\varepsilon^2
        }
        -\parentheses{1+\varepsilon}\dsmodulus^3
    }\\
    &\ge \frac{n}{2\dsmodulus^4}\parentheses{
        n\parentheses{
            \parentheses{\dsmodulus\varepsilon^2-1}\dsmodulus^2
            -\dsmodulus\parentheses{\dsmodulus-1}\varepsilon^2
        }
        -\parentheses{1+\varepsilon}\dsmodulus^3
    }\\
    &> 0,
\end{align*}
where the last inequality follows from Equation~\ref{eq:constraint2}.

Thus every committee with $1\le t\le\dsmodulus$ has strictly fewer \pairs edges than $X$, so $X$ is the unique \pairs maximizer.

For completeness, the same ledger identifies the opposite extreme. For fixed $t$, the \pairs score is nonincreasing in $\ell$, and $\ell\le t$: every uncovered small group is incident, through its $\dsk$ approved $X$-candidates, only to removed candidates, while each removed $X$-candidate is approved by exactly $\dsk$ groups. Hence the score of any committee with parameter $t$ is at least the score obtained by substituting $\ell=t$. This lower envelope is strictly decreasing in $t$. Indeed, since $n\varepsilon/\dsmodulus^2\in\N$ we have $n\varepsilon\ge\dsmodulus^2$, and a direct subtraction shows that its decrease from $t$ to $t+1$ is minimized at $t=\dsmodulus-1$ and is at least
\[
\frac{n}{2\dsmodulus^3}
\left[
 n\parentheses{\dsmodulus^2\varepsilon^2-\dsmodulus\varepsilon^2-\dsmodulus+\varepsilon^2}
 -\dsmodulus^2-\dsmodulus\varepsilon
\right]
>0,
\]
where the final inequality follows from Equation~\ref{eq:constraint2}. Consequently the unique \pairs-minimizing committee is obtained at $t=\ell=\dsmodulus$, namely $Y$.

Under $X$, all $\varepsilon n$ voters in the small set form one connected component, while every voter in the large set is isolated. Thus the largest connected component of the unique \pairs-optimal committee has size exactly $\varepsilon n<\delta n$. This proves the theorem.

%%%%%%%%%%%%%%%%%%%%%%%%%%%%%%%%%%%%%%%%%%%%%%%%%%%%%%%%%%%%%%%%%%%%%%%%

\section{CONS Sub-optimality Proofs}
\label{app:cons-subopt-proofs}

\subsection{Proof of CI Gateway Lemma}
\begin{proof}[Proof of Lemma~\ref{lem:ci-gateway}]
    Fix a \CIDomain{} candidate order and list the candidates of $\Committee$ with
nonempty support in that order as $c_1<\cdots<c_r$.  Since
$\CommitteeGraph{\Committee}$ is connected, every voter approves at least one
selected candidate.

First observe that the supports of consecutive selected candidates intersect.
Indeed, if the supports of $c_i$ and $c_{i+1}$ were disjoint, no voter could
approve a selected candidate weakly to the left of $c_i$ and another selected
candidate weakly to the right of $c_{i+1}$: by the \CIDomain{} property such a voter
would approve both $c_i$ and $c_{i+1}$.  The preferences graph would therefore split
between the selected candidates on the two sides, contradicting
connectedness.

Suppose a cut edge results by two voters that jointly approve a candidate $d$. First, candidates
belonging to $\Committee$ satisfy the claim by definition.
Now consider an arbitrary candidate $d \notin \braces{c_1,\ldots,c_r}$.  Suppose first that
$c_i<d<c_{i+1}$.  Every voter approving $d$ also approves at least one selected
candidate.  If that selected candidate lies weakly to the left of $c_i$, the
\CIDomain{} property forces the voter to approve $c_i$; if it lies weakly to the right
of $c_{i+1}$, the voter must approve $c_{i+1}$.  Hence the support of $d$ is
contained in the union of the supports of $c_i$ and $c_{i+1}$.

If neither $c_i$ nor $c_{i+1}$ crosses the cut, then, because their supports
intersect, both supports lie on the same side of the cut, so $d$ cannot create
a crossing edge.  If exactly one of them crosses, every crossing edge created
by $d$ has an endpoint in the support of that crossing selected candidate.  If
both cross, both endpoints lie in the union of two crossing selected supports.
Whenever a selected candidate crosses the cut, every voter in its support is
incident to an edge of $\Boundary[\CommitteeGraph{\Committee}]{S}$.  Thus in
all cases every crossing edge created by $d$ has an endpoint in
$R_S\cup R_{\bar S}$.

The same argument applies to candidates before $c_1$ or after $c_r$: their
support is contained in the support of the nearest selected candidate, and if
they cross the cut that selected candidate crosses it as well.
\end{proof}

%%%%%%%%%%%%%%%%%%%%%%%%%%%%%%%%%%%%%%%%%%%%%%%%%%%%%%%%%%%%%%%%%%%%%%%%

\subsection{Proof Details for the CI Worst Case Cheeger Ratio}
\begin{proof}
    For $\Committee_G$, consider the following six voter blocks:
\[
\begin{array}{c|c}
X_1 & \braces{a}\text{- and }\braces{a,b}\text{-voters}\\
X_2 & \braces{a,b,c}\text{-voters}\\
X_3 & \braces{b,c}\text{-voters}\\
X_4 & \braces{c,d}\text{-voters}\\
X_5 & \braces{c,d,e}\text{- and }\braces{d,e}\text{-voters}\\
X_6 & \braces{e}\text{-voter}.
\end{array}
\]
Their sizes are
\[
    t+2,\quad t,\quad t+1,\quad t,\quad t+1,\quad 1.
\]
The graph $\CommitteeGraph{\Committee_G}$ contains as a spanning subgraph the
blow-up of the path
\[
    X_1-X_2-X_3-X_4-X_5-X_6,
\]
where every consecutive pair of blocks is complete bipartite.  Denote this
spanning subgraph by $H$.  Since adding edges cannot decrease expansion,
\[
    \Cheeger{\CommitteeGraph{\Committee_G}}\ge\Cheeger{H}.
\]

Fix a nonempty $S\subseteq\VoterSet$ with $s=\abs{S}\le n/2$.  Put
$x_i=\abs{S\cap X_i}$ and, for $1\le i\le5$,
\[
    r_i=\min\braces{x_i,\abs{X_i}-x_i},
    \qquad
    R=\sum_{i=1}^5 r_i.
\]
For each of the first four interfaces $X_i-X_{i+1}$, its contribution $B_i$
to the boundary satisfies
\[
    B_i\ge t r_i
    \qquad\text{and}\qquad
    B_i\ge t r_{i+1},
\]
because the neighboring block has size at least $t$.  Hence
\[
    \abs{\Boundary[H]{S}}
    \ge
    \frac{t}{2}R.
\]
If $R\ge s/3$, then
\[
    \frac{\abs{\Boundary[H]{S}}}{s}\ge\frac{t}{6}.
\]

Suppose now that $R<s/3$.  If $s=1$, every vertex of $H$ has degree at least
$t$, so the expansion of $S$ is at least $t$.  Let $s\ge2$.  Among
$X_1,\ldots,X_5$, not all blocks can have at most half of their vertices in
$S$, since then $s\le R+1<s/3+1$, a contradiction.  Nor can all five blocks
have at least half of their vertices in $S$: writing
$M=\sum_{i=1}^5\abs{X_i}=n-1$, this would give
\[
    s\ge M-R>M-s/3,
\]
so $s>3M/4>n/2$, again a contradiction.  Therefore, along the path
$X_1,\ldots,X_5$ there is either a block split exactly in half or a pair of
adjacent blocks with opposite strict majorities.  In the former case an
adjacent interface contributes at least $t^2/2$ crossing edges; in the latter
case the interface between the opposite-majority blocks contributes at least
$t^2/4$.  Thus in either case
\[
    \abs{\Boundary[H]{S}}\ge\frac{t^2}{4}.
\]
Since $s\le n/2=(5t+5)/2$,
\[
    \frac{\abs{\Boundary[H]{S}}}{s}
    \ge
    \frac{t^2}{2n}
    =
    \frac{t^2}{10t+10}
    \ge
    \frac{t}{20}.
\]
We conclude that
\[
    \cheeger{\Committee_G,\Election}
    \ge
    \frac{t}{20}.
\]
Conversely, the single $\braces{e}$ voter has degree $t+1$ in
$\CommitteeGraph{\Committee_G}$, so
\[
    \cheeger{\Committee_G,\Election}\le t+1.
\]
Thus $\cheeger{\Committee_G,\Election}=\Theta(t)$.
\end{proof}

%%%%%%%%%%%%%%%%%%%%%%%%%%%%%%%%%%%%%%%%%%%%%%%%%%%%%%%%%%%%%%%%%%%%%%%

\section{Background and Proofs for Information Diffusion Section}
\label{app:diffusion-background}

For a Markov chain $P$ with stationary distribution $\pi$ and a non-trivial cut $S\mid\bar S$, the \emph{conductance} of $S$ is defined as
\[
    \Phi\parentheses{S} = \frac{\sum_{i\in S,j\in\bar S}\pi_i P_{ij}}{\min\braces{\pi\parentheses{S},\pi\parentheses{\bar S}}}
\]
where $\pi\parentheses{.}$ is the total sum of weights in the stationary distribution of nodes in the subset. Conductance of $P$ is then defined as
\[
    \Phi(P):=\min_{\emptyset\ne S\subsetneq V} \Phi\parentheses{S}.
\]
If $P$ is symmetric and has nonnegative eigenvalues, spectral theory and the Cheeger inequalities~\cite{cheeger1970lower,alon1985lambda1} bound the mixing time as a function of conductance, yielding
\[
\Omega\parentheses{\frac{1}{\Phi\parentheses{P}}} \le t_{mix} \le O\parentheses{\frac{1}{\Phi\parentheses{P}^2}\log\frac{1}{\pi_{min}}}
\]
With $\pi_{min}$ the minimal entry in the stationary distribution.

\paragraph{Mixing time.}
We use standard terminology and results from finite Markov-chain theory; unless stated otherwise, the background in this subsection follows ~\citet{levin2026markov}.\rmr{wrong year. you have to be really careful with references and check them manually}\gbs{I did. It's a new edition:)}
The \emph{mixing time} of a Markov process $t_{mix}$ is the number of steps needed for a Markov chain to approach its stationary distribution within a fixed convention accuracy threshold, regardless of its initial state. A Markov chain is guaranteed to converge to a stationary distribution if $P$ is aperiodic and irreducible. For a Markov chain $P$ with stationary distribution $\pi$ and a non-trivial cut $S\mid\bar S$, the \emph{conductance} of $S$ is defined as
\[
    \Phi\parentheses{S} = \frac{\sum_{i\in S,j\in\bar S}\pi_i P_{ij}}{\min\braces{\pi\parentheses{S},\pi\parentheses{\bar S}}}
\]
where $\pi\parentheses{.}$ is the total sum of weights in the stationary distribution of nodes in the subset. Conductance of $P$ is then defined as
\[
    \Phi(P):=\min_{\emptyset\ne S\subsetneq V} \Phi\parentheses{S}.
\]
If $P$ is symmetric and has nonnegative eigenvalues, spectral theory and the Cheeger inequalities~\cite{cheeger1970lower,alon1985lambda1} bound the mixing time as a function of conductance, yielding
\[
\Omega\parentheses{\frac{1}{\Phi\parentheses{P}}} \le t_{mix} \le O\parentheses{\frac{1}{\Phi\parentheses{P}^2}\log\frac{1}{\pi_{min}}}
\]
With $\pi_{min}$ the minimal entry in the stationary distribution.

\paragraph{DeGroot and consensus.}
In a classic DeGroot model, interactions are described by an influence matrix $P$ in which $P_{ij}$ is the relative influence that agent $i$ assigns to agent $j$'s opinion~\cite{demarzo2003persuasion}. The weights are relative in the sense that, for each agent, the total influence assigned across all agents is normalized to one; equivalently, \(P\) is row-stochastic. Let \(x(t)\) denote the vector of agents' opinions at time \(t\). The DeGroot update is
\[
    x(t+1)=Px(t).
\]
The same matrix can be viewed as the transition matrix of a Markov chain, which allows the convergence of the averaging process to be studied using standard spectral and mixing tools.

It is standard in DeGroot-type opinion-dynamics models to interpret the diagonal entry \(P_{ii}\) as agent \(i\)'s self-reliance, and positive self-weights \(P_{ii}>0\) are a common modeling assumption~\cite{jia2015opinion}.\rmr{they also use zero self-weight} If we examine each connected component individually, $P$ reduced to the nodes of the connected component is irreducible, and the positive self-weights make it aperiodic. Thus, convergence of each component separately is guaranteed. Moreover, the opinion value of each agent converges to the same consensus value which is a weighted average of the agents initial opinions~\cite{degroot1974reaching,berger1981necessary,olfati2004consensus}.
Under the DeGroot interpretation, convergence of the Markov chain to its stationary distribution corresponds to convergence of opinions to a steady state, namely consensus. Accordingly, the Markov-chain mixing time coincides, up to the chosen accuracy convention, with the time required for the DeGroot process to reach approximate consensus for normalized initial opinions.

%%%%%%%%%%%%%%%%%%%%%%%%%%%%%%%%%%%%%%%%%%%%%%%%%%%%%%%%%%%%%%%%%%%%%%%

\subsection{Proof of Theorem~\ref{thm:diffusion-disconnected-divergence}}
\begin{proof}
    Any restriction of \(\DiffusionMatrix[\beta]{\Committee}\) to the connected component yields an aperiodic irreducible matrix, guaranteeing convergence. Since the matrix is symmetric, it is doubly stochastic. A standard result for finite Markov chains is that every doubly stochastic transition matrix admits the uniform stationary distribution. Thus, for a connected component $B\subseteq\VoterSet$
    \[
        \lim_{t\to\infty}x_i(t)
        =
        \frac{1}{\abs{B}}\sum_{j\in B}x_j(0)
        \qquad
        \text{for every }i\in B,
    \]
    with $x(0)$ the initial opinions of voters. In particular, if $\Committee$ is connecting, $B=\VoterSet$.
\end{proof}

\subsection{Proof of Theorem~\ref{thm:diffusion-bounded-consensus}}

\begin{proof}
    As discussed above, the stationary distribution is uniform. Thus, the conductance of $S\subseteq\VoterSet$ with $0<\abs{S}\le n/2$ is
\[
    \SetConductance[\beta]{\Committee}{S}
    =
    \frac{
        \sum_{i\in S,\,j\notin S}
        \frac1n \bigl(\DiffusionMatrix[\beta]{\Committee}\bigr)_{ij}
    }{
        \abs{S}/n
    }
    =
    \frac{\beta}{n-1}
    \frac{\abs{\Boundary[G_{\Committee}]{S}}}{\abs{S}}
\]
Consequently,
\[
    \ChainConductance{\DiffusionMatrix[\beta]{\Committee}}
    =
    \frac{\beta}{n-1}\Cheeger{G_{\Committee}}
\]
In particular, for $\beta=1/2$
\[
    \ChainConductance{\DiffusionMatrix[1/2]{\Committee}}
    =
    \frac{\Cheeger{G_{\Committee}}}{2(n-1)}.
\]
Since the lazy process with $\beta=1/2$ admits nonnegative eigenvalues, the conductance bounds imply the requested bounds.
\end{proof}

\section{Omitted Proofs for the Complexity Section}
\label{app:complexity-proofs}

\subsection{Fixed-committee CHEEGER evaluation}
\label{app:fixed-committee-hardness}

\begin{proof}[Proof of Theorem~\ref{thm:fixed_committee_cheeger_npcomplete}]
Membership follows from a cut certificate: a nonempty set $S$ with
$\abs S\le\abs{\VoterSet}/2$ certifies the answer whenever
\[
    \frac{
      \abs{\Boundary[\CommitteeGraph[\Election]{\Committee}]{S}}
    }{
      \abs S
    }
    \le\Threshold.
\]

For hardness, reduce from the Cheeger decision problem on a general graph~\cite{leighton1999multicommodity,garey1974some}
$G=(V,E)$.  Set
\[
    \VoterSet=V,
    \qquad
    \CandidateSet=E,
\]
and for every $v\in V$ let $\ApprovalBallot{v}$ consist of the edges of
$G$ incident to $v$.  Set $\Committee=E$ and
$\CommitteeSize=\abs E$.  Two voters jointly approve a committee member
exactly when the corresponding vertices are adjacent in $G$, and hence
\[
    \CommitteeGraph[\Election]{\Committee}=G.
\]
The construction has $\abs V$ voters, $\abs E$ candidates, and $2\abs E$
approval incidences, so it is polynomial.  Therefore
\[
    \cheeger{\Committee,\Election}\le\Threshold
    \quad\Longleftrightarrow\quad
    \Cheeger{G}\le\Threshold.
\]
\end{proof}

\subsection{NP-hardness of exact-size CHEEGER selection on \VIDomain{}}
\label{app:vi-hardness}

\subsection{Problem Statement}

We study the computational complexity of the following decision problem.

\begin{computationalproblem}{VI-Exact-Cheeger}
    \item[\textbf{Input:}]
    \VIDomain{} election $\Election=(\CandidateSet,\VoterSet,\ApprovalProfile,\CommitteeSize)$ and a threshold \(\Threshold\in\Q^+\).

    \item[\textbf{Question:}]
    Does a committee
    \(\Committee\in\CommitteeFamily\) satisfying $\cheeger{\Committee,\Election}\ge \Threshold$ exists?
\end{computationalproblem}

Equivalently, the question is whether there exists
\(\Committee\subseteq\CandidateSet\) with
\(\abs{\Committee}=\CommitteeSize\) and
\[
\Cheeger{\CommitteeGraph[\Election]{\Committee}}\ge\Threshold.
\]

For the hardness proof, we reduce from the following unary variant of
\textsc{3-Partition}.

\begin{computationalproblem}{Strict-Unary-3-Partition}
    \item[\textbf{Input:}]
    Positive integers $\VIHSourceWeight{1},\ldots,\VIHSourceWeight{\VIHSourceItems},\VIHSourceTarget$,
    encoded in unary, where
    \[
    \VIHSourceItems=3\VIHSourceBins,
    \qquad
    \sum_{i=1}^{\VIHSourceItems}\VIHSourceWeight{i}
    =
    \VIHSourceBins\VIHSourceTarget,
    \]
    and
    \[
    \frac{\VIHSourceTarget}{4}
    <
    \VIHSourceWeight{i}
    <
    \frac{\VIHSourceTarget}{2}
    \qquad
    \text{for every }i.
    \]

    \item[\textbf{Question:}]
    Is there a partition of the items into \(\VIHSourceBins\) pairwise disjoint sets $\VIHSourceTriple{1},\ldots,\VIHSourceTriple{\VIHSourceBins}$ such that for every $j$,
    \[
    \sum_{i\in\VIHSourceTriple{j}}
    \VIHSourceWeight{i}
    =
    \VIHSourceTarget?
    \]
\end{computationalproblem}

\StrictUnaryThreePartition is known to be NP-complete~\cite{garey2002computers}.

We prove NP-hardness by a polynomial-time reduction
\[
\textsc{Strict-Unary-3-Partition}
\le_p
\textsc{VI-Exact-Cheeger}.
\]
More precisely, from a source instance \(I\) we construct an instance
\((\VIHConstructedElection,\Threshold)\) such that
\[
I\text{ is a YES-instance}
\quad\Longleftrightarrow\quad
(\VIHConstructedElection,\Threshold)
\text{ is a YES-instance of \textsc{VI-Exact-Cheeger}}.
\]

\begin{theorem}
\VICheegerExact{} is NP-hard, even when every candidate support is
nonempty and the candidate supports are pairwise distinct proper intervals.
\end{theorem}

\subsection{Reduction idea}

The construction outputs an election $\VIHConstructedElection$; unless stated otherwise, graph notation below refers to this constructed instance. It encodes a solution as a sequence of cumulative sets of items referred to as \emph{stages}, with the interpretation that stage $h$ encodes all items assigned to the first $h$ bins of the solution. For each $h<\VIHSourceBins$, stage $h$ is represented by two \emph{variable phases}, $2h$ and $2h+1$, which respectively impose a lower and an upper bound on the total weight of the items in the stage. The final stage $h=\VIHSourceBins$ is represented only by the final even phase $2\VIHSourceBins$, as its upper weight bound follows from the total weight of all items.

For each phase $p$ and item $i$, there are two alternative candidates $\VIHBranchA{p}{i}$ and $\VIHBranchB{p}{i}$. The construction will force exactly one of them to be selected. We write
\[
\VIHChoice{p}{i}=
\begin{cases}
1,&\VIHBranchA{p}{i}\text{ is selected},\\
0,&\VIHBranchB{p}{i}\text{ is selected},
\end{cases}
\]

The intended interpretation is that $\VIHChoice{p}{i}=1$ once item $i$ has entered the cumulative collection of completed 3-PARTITION bins. This monotone property is enforced by phase and item specific markers, specifically selected prefix cuts, in which if candidate $\VIHBranchA{p}{i}$ is selected but $\VIHBranchA{p+1}{i}$ is not, then the expansion drops below the threshold. Thus any committee meeting the threshold ensures $\VIHChoice{p}{i}=1 \rightarrow \VIHChoice{p'}{i}=1$ for every subsequent phase $p'$.

Additionally, for each phase $p$ and item $i$, two auxiliary weighting candidates, an anchor $\VIHItemAnchor{p}{i}$ and a weight-dependent component $\VIHMergeCandidate{p}{i}$ are introduced. Together with the two alternative candidates, they create $a_i$-sized multiplier for $\VIHChoice{p}{i}$. This is achieved using a weight probe, a single phase specific selected prefix, used to measure the cumulative weight effect of all items. At the weight probe, each individual item's weighting candidates create an $a_i$ difference in boundary by choosing $\VIHBranchA{p}{i}$ relative to choosing $\VIHBranchB{p}{i}$. Consequently, $\VIHEncodedWeight{p}=\sum_i \VIHSourceWeight{i} \VIHChoice{p}{i}$ is the cumulative weight encoded in phase $p$.

The two phases associated with a stage encode this same cumulative weight with opposite signs: in an even phase, selecting $\VIHBranchA{p}{i}$ contributes $+\VIHSourceWeight{i}$ to the relevant boundary expression relative to selecting $\VIHBranchB{p}{i}$, while in an odd phase it contributes $-\VIHSourceWeight{i}$. Hence the committee-dependent contribution of phase $p$ is proportional to $(-1)^p\VIHEncodedWeight{p}$. 

Each phase also includes a small set of calibration candidates with the intended purpose of comparing $\VIHEncodedWeight{p}$ against the desired $h\VIHSourceTarget$ cumulative weight at stage $h$. Even and odd phases impose opposite inequalities on the cumulative weight represented by that stage. The calibration candidates cause the baseline boundary of the weight probe to correspond with the target cumulative weight. The alternating sign then makes weights below the target fail in the even phase and weights above the target fail in the odd phase, so the two phases jointly impose both a lower and an upper bound on the cumulative weight represented by the stage. Therefore, any committee that guarantees the threshold must encode cumulative weight exactly $h\VIHSourceTarget$ at stage $h$. 

Because
\[
\frac{\VIHSourceTarget}{4}
<
\VIHSourceWeight{i}
<
\frac{\VIHSourceTarget}{2},
\]
every set of total weight \(\VIHSourceTarget\) necessarily contains exactly
three items.
Consequently, the differences between stages form a valid 3-partition.

The remaining candidates have three auxiliary purposes: they calibrate all diagnostic cuts (i.e., monotonicity markers and weight probes) to one common threshold, force every feasible committee into the intended $\VIHBranchALabel/\VIHBranchBLabel$ form, and provide enough surplus to make all nondiagnostic cuts safe, i.e. have expansion above the threshold. The construction is analyzed first through prefix cuts of the voter order, which are where the monotonicity markers and weight probes are placed. This does not assume that a minimum-expansion cut is a prefix: after establishing the required bounds for all prefixes, we separately show that every arbitrary cut is at least as safe, using the interval structure together with the large head and tail cliques.

\subsection{Construction}

\paragraph{The Neutral Phase.}
For the analysis, it is useful to first consider a simplified version of a variable phase, which we call a \emph{neutral phase}. A neutral phase is constructed from the same calibration, weighting, and alternative candidates, with additional scaffolding candidates that control fixed numerical values used in the construction, and serves as a common baseline against which the actual phases will later be compared.

A phase should not be thought of as an isolated gadget whose candidates are contained within it. Rather, $\VIHPhaseSpan$ specifies a fixed-size window in the voter order, and candidate intervals typically extend from one such window into the next. Let $\VIHPhaseOrigin{p}$ denote the beginning of phase $p$. We describe a candidate in phase $p$ by a local pair $(a,b)$, meaning that its approval interval is
\[
\bigl[
\VIHPhaseOrigin{p}+a+1,\,
\VIHPhaseOrigin{p+1}+b
\bigr].
\]
Thus $a$ specifies the position of its left endpoint relative to the beginning of the current phase, whereas $b$ specifies the position of its right endpoint relative to the beginning of the following phase. Specifically, set
\[
\VIHPhaseOrigin{1}=(\VIHBoundaryTarget-1)\VIHDenominator+1,\qquad \VIHPhaseOrigin{0}=\VIHPhaseOrigin{1}-\VIHPhaseSpan
\]
for a parameter $\VIHBoundaryTarget$ that we calculate later. Then for each phase $p$ $\VIHPhaseOrigin{p}=\VIHPhaseOrigin{0}+p\VIHPhaseSpan$ is the start of phase $p$.

Since the intervals of voters approving each candidate extend between phases, the geometry of a neutral phase is easiest to describe through the positions at which candidates begin. Set
\[
\VIHSourceItems=3\VIHSourceBins,\qquad \VIHNumPhases=2\VIHSourceBins+1,\qquad \VIHNumPairs=\VIHNumPhases\VIHSourceItems,\qquad \VIHTotalWeight=\VIHSourceBins\VIHSourceTarget.
\]
And define
\[
\VIHMergeScale=100(\VIHNumPairs+3),\qquad \VIHPerturbBudget=(\VIHMergeScale+1)\VIHTotalWeight,\qquad \VIHSlack=10\VIHNumPhases\VIHPerturbBudget+10,
\]
\[
\VIHItemScale=50(\VIHSlack+\VIHNumPairs+\VIHTotalWeight+\VIHMergeScale+1),\qquad \VIHBranchGap=2\VIHItemScale,
\]
\[
\VIHScaffoldScale=100(\VIHNumPairs+\VIHNumPhases+10)\VIHBranchGap,\qquad \VIHPhaseSpan=12\VIHScaffoldScale+4\VIHSourceItems\VIHItemScale.
\]

As discussed above each phase has size $\VIHPhaseSpan$. The first $8H$ positions are used by the scaffold. The item gadgets then appear in order, with
\[
\VIHItemOrigin{i}=8H+4\VIHItemScale(i-1),
\]
so the starting patterns of consecutive items are spaced $4\VIHItemScale$ apart. Relative to $\VIHItemOrigin{i}$, the two alternatives begin at
\[
\VIHItemOrigin{i}+\VIHItemScale-1
\qquad\text{and}\qquad
\VIHItemOrigin{i}+\VIHItemScale,
\]
the auxiliary weighting candidate $\VIHMergeCandidate{p}{i}$ begins at
\[
\VIHItemOrigin{i}+3\VIHItemScale
\]
in the neutral phase, and the anchor $\VIHItemAnchor{p}{i}$ begins at
\[
\VIHItemOrigin{i}+4\VIHItemScale.
\]

\begin{figure}[ht]
\centering
\begin{tikzpicture}[x=0.95cm,y=0.85cm,>=Latex,
  cand/.style={line width=1.1pt},
  every node/.style={font=\small}
]

% two phases baseline
\draw[thick] (0,0) -- (13.5,0);
\draw[densely dashed] (6.2,-1.2) -- (6.2,3.1);
\node[below] at (3.1,0) {phase $p$};
\node[below] at (9.85,0) {phase $p{+}1$};

% current item window in phase p
\draw[fill=black!6] (1.2,0.15) rectangle (5.2,0.85);
\node at (3.2,0.5) {item window for $i$};
\draw[decorate,decoration={brace,mirror,amplitude=5pt}] (1.2,-0.15) -- (5.2,-0.15)
  node[midway,below=7pt] {$4v$};

% corresponding window in next phase
\draw[fill=white] (7.4,0.15) rectangle (11.4,0.85);
\node[font=\scriptsize] at (9.4,0.5) {next phase};

% xi labels
\node[below=2pt] at (1.2,0) {$\xi_i$};
\node[below=2pt] at (2.2,0) {$\xi_i+v$};
\node[below=2pt] at (4.2,0) {$\xi_i+3v$};
\node[below=2pt] at (5.2,0) {$\xi_i+4v$};

% candidates with black-and-white styles
% A solid
\draw[cand] (2.0,2.6) -- (8.4,2.6);
\fill (2.0,2.6) circle (1.8pt);
\node[left] at (2.0,2.6) {$A_{p,i}$};
\node[font=\scriptsize,above] at (2.0,2.6) {$\xi_i+v-1$};
\node[font=\scriptsize,above] at (8.4,2.6) {$X_{p+1}+\xi_i+v$};

% B dashed
\draw[cand,dashed] (2.2,2.0) -- (8.6,2.0);
\fill (2.2,2.0) circle (1.8pt);
\node[left] at (2.2,2.0) {$B_{p,i}$};
\node[font=\scriptsize,above] at (2.2,2.0) {$\xi_i+v$};
\node[font=\scriptsize,above] at (8.6,2.0) {$X_{p+1}+\xi_i+v+1$};

% e dash dot
\draw[cand,dash dot] (4.2,1.4) -- (8.8,1.4);
\fill (4.2,1.4) circle (1.8pt);
\node[left] at (4.2,1.4) {$e_{p,i}$};
\node[font=\scriptsize,above] at (4.2,1.4) {$\xi_i+3v$};
\node[font=\scriptsize,above] at (8.8,1.4) {$X_{p+1}+\xi_i+v+1+J$};

% shifted start arrow for e
\draw[->] (3.6,1.1) -- (4.15,1.32);
\node[font=\scriptsize,anchor=east] at (3.55,1.05)
  {phase-specific shift by $\sigma_p a_i$};

% g dotted
\draw[cand,dotted] (5.2,0.8) -- (10.4,0.8);
\fill (5.2,0.8) circle (1.8pt);
\node[left] at (5.2,0.8) {$g_{p,i}$};
\node[font=\scriptsize,above] at (5.2,0.8) {$\xi_i+4v$};
\node[font=\scriptsize,above] at (10.4,0.8) {$X_{p+1}+\xi_i+3v$};

% local markers in window
\foreach \xx in {2.2,4.2,5.2}{
  \draw[densely dotted] (\xx,0) -- (\xx,3.0);
}
\node[font=\scriptsize] at (3.2,3.0) {left endpoints in phase $p$};
\node[font=\scriptsize] at (9.3,3.0) {right endpoints in phase $p{+}1$};

% legend
\node[draw, align=left, anchor=west, font=\scriptsize] at (10.6,2.3)
{line styles:\
solid $=A_{p,i}$\\
dashed $=B_{p,i}$\\
dash-dot $=e_{p,i}$\\
dotted $=g_{p,i}$};

\end{tikzpicture}
\caption{The item gadget for item $i$. The left endpoints of the four candidates lie in the $4v$-wide item window of phase $p$, while their right endpoints lie in phase $p+1$. In the neutral template, $e_{p,i}$ starts at $\xi_i+3v$; in the actual construction this start is shifted by $\sigma_p a_i$.}
\end{figure}
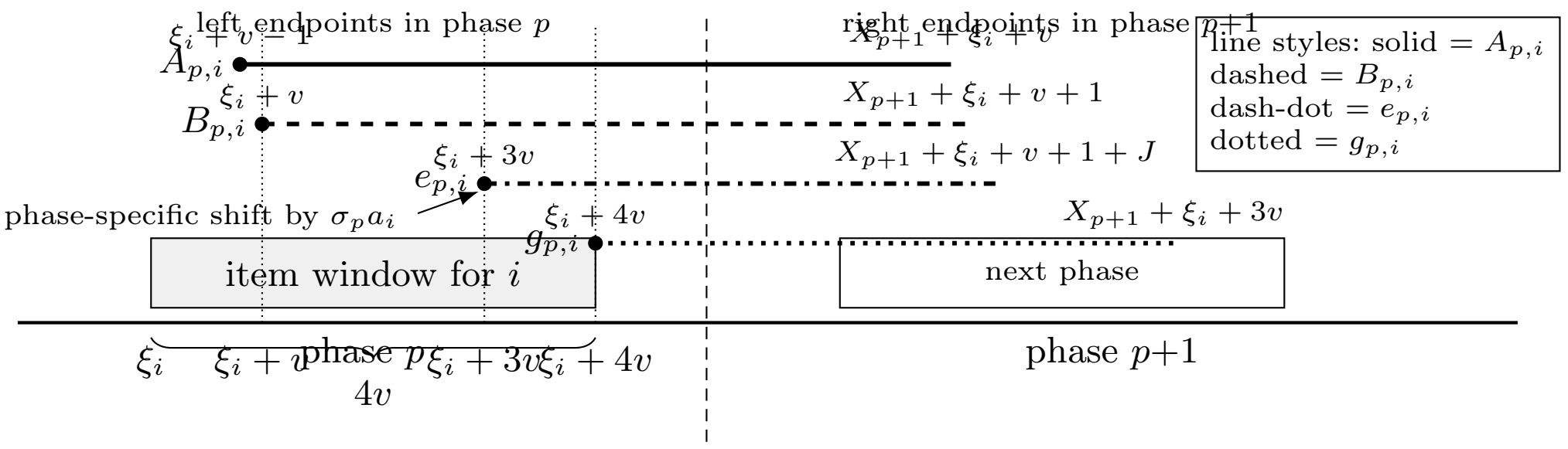

The neutral scaffold consists of seven fixed candidates, whose local coordinates are
\[
\begin{array}{c|cc}
& a & b\\ \hline
c_0 & 0 & -2H\\
c_1 & H & -H\\
c_2 & H+1 & 0\\
c_3 & 3H & 5H\\
c_4 & 4H & 6H\\
c_5 & 4H+1 & 7H\\
g_0 & 8H & 8H-v.
\end{array}
\]
Here, as above, $(a,b)$ means that the candidate interval starts at position $a$ relative to the current phase and ends at position $b$ relative to the following phase. Among these scaffold candidates, $c_1$ and $c_4$ are the \emph{calibration candidates}.

Note that these are only the starting positions: the corresponding candidate intervals extend into the following phase according to their second local coordinate.

We concatenate $2\VIHSourceBins+1$ copies of the neutral phase template, each spanning $\VIHPhaseSpan$ positions in the voter order. These copies form the variable phases $p=0,\ldots,2\VIHSourceBins$. We then apply the phase-specific modifications described below.

\paragraph{Phase-Specific Modifications}
Actual phases are obtained from the neutral template by introducing the
phase-specific information. For each variable phase $p$, let
\[
\VIHPhaseSign{p}=(-1)^p.
\]
The weighting candidate $\VIHMergeCandidate{p}{i}$ for item $i$ is moved from
its neutral starting position $\VIHItemOrigin{i}+3\VIHItemScale$ to
\[
\VIHItemOrigin{i}+3\VIHItemScale
-\VIHPhaseSign{p}\VIHSourceWeight{i}.
\]
Thus, the weight candidate of each item has offset of $\VIHSourceWeight{i}$ from the neutral phase's position. $\VIHPhaseSign{p}$ changes the orientation of the offset in even and odd phases.

In addition, the two calibration candidates in phase $p$ are similarly modified. Let
\[
\VIHPhaseTarget{p}=\left\lfloor\frac p2\right\rfloor \VIHSourceTarget.
\]
The calibration value is set to be
\[
\VIHCalibration{p}=-p\VIHSlack-
\begin{cases}
\VIHMergeScale\VIHTotalWeight+\VIHPhaseTarget{p},&p\text{ odd},\\
0,&p\text{ even}.
\end{cases}
\] 
Then the right endpoint of \(c_1\) is shifted by \(+\VIHCalibration{p}\), while the right endpoint of \(c_4\) is shifted by \(-\VIHCalibration{p}\).

\begin{figure}[ht]
\centering
\begin{tikzpicture}[x=0.72cm,y=0.8cm,>=Latex,every node/.style={font=\small}]

% Phase outline
\draw[thick] (0,0) rectangle (16,1.4);
\node[above] at (8,1.4) {phase $p$ of length $\Omega$};

% Scaffold region
\draw[thick] (0,0) rectangle (4.8,1.4);
\foreach \x in {0.2,0.55,...,4.6} {
  \draw[thin] (\x,0) -- ++(0.7,1.4);
}
\node[fill=white,inner sep=1pt] at (2.4,0.95) {scaffold};
\node[fill=white,inner sep=1pt,font=\scriptsize] at (2.4,0.45)
  {$c_0,c_1,c_2,c_3,c_4,c_5,g_0$};

% Item windows
\draw (4.8,0) rectangle (6.8,1.4);
\draw (6.9,0) rectangle (8.9,1.4);
\draw (11.1,0) rectangle (13.1,1.4);
\draw (13.2,0) rectangle (15.2,1.4);

\node at (5.8,0.95) {item $1$};
\node[font=\scriptsize] at (5.8,0.45) {width $4v$};

\node at (7.9,0.95) {item $2$};
\node[font=\scriptsize] at (7.9,0.45) {width $4v$};

\node at (10.0,0.70) {$\cdots$};

\node at (12.1,0.95) {item $i$};
\node[font=\scriptsize] at (12.1,0.45) {width $4v$};

\node at (14.2,0.95) {item $q$};
\node[font=\scriptsize] at (14.2,0.45) {width $4v$};

% Braces
\draw[decorate,decoration={brace,mirror,amplitude=5pt}]
  (0,-0.15) -- (4.8,-0.15)
  node[midway,below=7pt] {$8H$};

\draw[decorate,decoration={brace,mirror,amplitude=5pt}]
  (4.8,-0.15) -- (16,-0.15)
  node[midway,below=7pt] {$4qv$};

% Phase endpoints
\node[below=2pt] at (0,0) {$X_p$};
\node[below=2pt] at (16,0) {$X_{p+1}$};

\end{tikzpicture}
\caption{A single phase. The phase consists of a scaffold block followed by $q$ item windows.}
\end{figure}
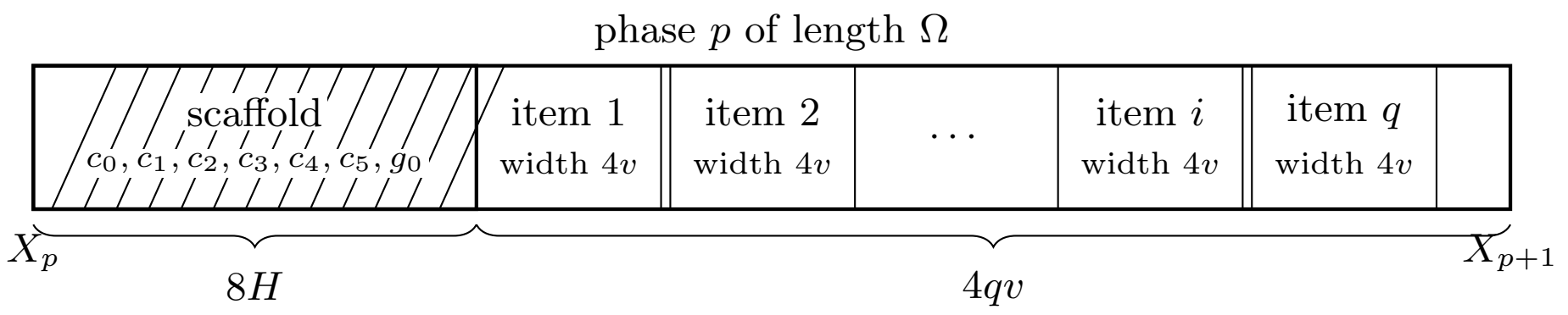

There are two endpoint exceptions to the common phase template. Phase $0$
omits the first three scaffold candidates
$\VIHScaffoldCandidate{0},\VIHScaffoldCandidate{1},
\VIHScaffoldCandidate{2}$. After the last variable phase, the construction
contains one final scaffold-only phase: it contains the scaffold candidates
but no item gadgets.

The final phase-specific intervals are therefore summarized below.
\[
\begin{array}{c|cc}
 &a&b\\ \hline
\VIHScaffoldCandidate{0}&0&-2\VIHScaffoldScale\\
\VIHScaffoldCandidate{1}&\VIHScaffoldScale&-\VIHScaffoldScale+\VIHCalibration{p}\\
\VIHScaffoldCandidate{2}&\VIHScaffoldScale+1&0\\
\VIHScaffoldCandidate{3}&3\VIHScaffoldScale&5\VIHScaffoldScale\\
\VIHScaffoldCandidate{4}&4\VIHScaffoldScale&6\VIHScaffoldScale-\VIHCalibration{p}\\
\VIHScaffoldCandidate{5}&4\VIHScaffoldScale+1&7\VIHScaffoldScale\\
\VIHScaffoldAnchor&8\VIHScaffoldScale&8\VIHScaffoldScale-\VIHItemScale.
\end{array}
\tag{scaffold}
\]

\[
\begin{array}{c|cc}
&a&b\\ \hline
\VIHBranchA{p}{i}&\VIHItemOrigin{i}+\VIHItemScale-1&\VIHItemOrigin{i}+\VIHItemScale\\
\VIHBranchB{p}{i}&\VIHItemOrigin{i}+\VIHItemScale&\VIHItemOrigin{i}+\VIHItemScale+1\\
\VIHMergeCandidate{p}{i}&\VIHItemOrigin{i}+3\VIHItemScale-\VIHPhaseSign{p}\VIHSourceWeight{i}&\VIHItemOrigin{i}+\VIHItemScale+1+\VIHMergeScale\\
\VIHItemAnchor{p}{i}&\VIHItemOrigin{i}+4\VIHItemScale&\VIHItemOrigin{i}+3\VIHItemScale.
\end{array}
\tag{items}
\]

Finally, we embed the phased construction inside a larger voter order. We call the area that contains the concatenated phases the active region.

The active region begin in $\VIHPhaseOrigin{1}$ and we denote by $\VIHActiveEnd=\VIHPhaseOrigin{\VIHNumPhases}+8\VIHScaffoldScale+\VIHItemScale-1$ the right endpoint of the active region; the total number of voters is denoted by $n$. Their exact values are chosen later so that every prefix through the active region lies on the smaller side of the Cheeger cut and so that there is ample padding on both sides.

We use this padding to add two fixed candidates. The head candidate is approved by voters $[1,\VIHPhaseOrigin{1}]$, while the tail candidate is approved by voters $[\VIHActiveEnd+1,n]$. Thus each end of the construction forms a large clique in the committee graph. These cliques make prefixes lying sufficiently far outside the phased construction automatically safe, allowing the remainder of the proof to focus on cuts through the active region.

\begin{figure}[ht]
\centering
\begin{tikzpicture}[x=0.7cm,y=0.8cm,>=Latex,
  every node/.style={font=\small}
]

% global baseline
\draw[thick] (0,0) -- (17.2,0);

% head with open-ended continuation
\draw[pattern=north east lines, pattern color=black] (0.9,0.25) rectangle (2.6,1.15);
\draw[thick,->] (0.9,0.70) -- (0.1,0.70);
\node at (1.75,0.7) {head};
\node[below=2pt] at (0.9,0) {$1$};
\node[below=2pt] at (2.6,0) {$X_1$};

% active region
\draw[fill=black!6] (2.6,0.25) rectangle (14.6,1.15);
\node at (10.6,1.45) {active region};
\node[below=2pt] at (14.6,0) {$\eta$};

% tail with open-ended continuation
\draw[pattern=north east lines, pattern color=black] (14.6,0.25) rectangle (16.3,1.15);
\draw[thick,->] (16.3,0.70) -- (17.1,0.70);
\node at (15.45,0.7) {tail};
\node[below=2pt] at (16.3,0) {$n$};

% phases inside active region
\foreach \i/\name in {0/0,1/1,2/{2h},3/{2h{+}1},4/{2h{+}2},5/{2h{+}3},6/{2s}}{
  \pgfmathsetmacro{\xleft}{2.85 + 1.62*\i}
  \draw[fill=white] (\xleft,0.38) rectangle ++(1.40,0.64);
  \node[font=\scriptsize] at (\xleft+0.70,0.70) {phase $\name$};
}

% ellipses
\node at (4.95,0.70) {$\cdots$};
\node at (12.15,0.70) {$\cdots$};

% dashed stage boxes
\draw[dashed, very thick, rounded corners] (5.97,0.18) rectangle (9.22,1.22);
\node[below=1pt] at (7.42,0) {stage $h$};

\end{tikzpicture}
\caption{The overall active region. The head and tail indicate padded cliques that continue beyond the active construction. Inside the active region, the construction concatenates the variable phases, and each stage is represented by two consecutive phases.}
\end{figure}
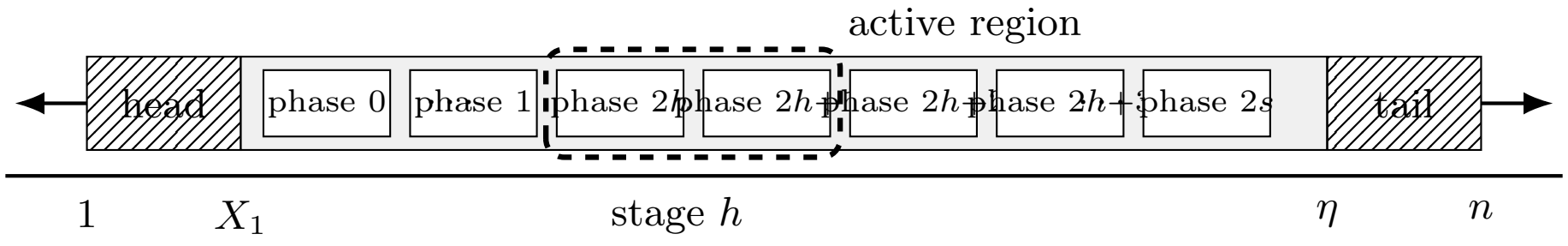

Finally, we set the committee size $\CommitteeSize=\VIHNumPhases(3\VIHSourceItems+7)+6$
and the threshold $\Threshold=\frac1\VIHDenominator$ for $\VIHDenominator=10(\VIHNumPhases+3)\VIHPhaseSpan$.

\subsection{Proof- general machinery}

\paragraph{A common threshold for all active prefixes}

The active region satisfies
\[
\VIHActiveEnd-\VIHPhaseOrigin{1}+1<\VIHDenominator.
\]
Since
\[
\VIHPhaseOrigin{1}=(\VIHBoundaryTarget-1)\VIHDenominator+1,
\]
every integer $t\in[\VIHPhaseOrigin{1},\VIHActiveEnd]$ satisfies
\[
(\VIHBoundaryTarget-1)\VIHDenominator<t<\VIHBoundaryTarget\VIHDenominator
\]
and lies below $n/2$. The Cheeger requirement at this prefix is
\[
\abs{\Boundary[\CommitteeGraph{\Committee}]{\VIHPrefix{t}}}\ge\frac{t}{\VIHDenominator}.
\]

$\abs{\Boundary[\CommitteeGraph{\Committee}]{\VIHPrefix{t}}}$ is an integer, therefore it must be that $\abs{\Boundary[\CommitteeGraph{\Committee}]{\VIHPrefix{t}}}\ge \VIHBoundaryTarget$.

Thus, for every prefix cut we only need to lower bound the boundary by $\VIHBoundaryTarget$ to show its expansion meets the threshold.

For completeness, every \emph{nontrivial} prefix outside the active window is automatically safe for the intended committee as well. If
\[
1\le t<\VIHPhaseOrigin{1},
\]
the head clique contributes
\[
t(\VIHPhaseOrigin{1}-t)\ge t,
\]
so in particular
\[
\abs{\Boundary[\CommitteeGraph{\Committee}]{\VIHPrefix{t}}}\ge\frac{\min\braces{t,n-t}}{\VIHDenominator}.
\]
If
\[
\VIHActiveEnd<t<n,
\]
the tail clique contributes
\[
(t-\VIHActiveEnd)(n-t)\ge n-t,
\]
and therefore again
\[
\abs{\Boundary[\CommitteeGraph{\Committee}]{\VIHPrefix{t}}}\ge\frac{\min\braces{t,n-t}}{\VIHDenominator}.
\tag{11}
\]

\paragraph{Signed contribution}

%The main reason for introducing the neutral phase is a simple invariant concerning boundaries of prefix cut, cuts in which the set of nodes in both sides is a continuous segment of voters.

Fix a committee $\Committee$ and a prefix $\VIHPrefix{t}=\{1,\ldots,t\}$. For a voter $x\le t$, let $\VIHRightReach{\Committee}{x}$ be the rightmost voter that shares a selected candidate with $x$. Since every selected candidate's support is an interval of voters, $x$ is adjacent to every voter between $x$ and $\VIHRightReach{\Committee}{x}$. Hence the number of boundary edges incident to $x$ and crossing $\VIHPrefix{t}$, to which we also call the contribution of $x$, is
\[
\max\{0,\VIHRightReach{\Committee}{x}-t\}.
\]

Therefore
\[
\abs{\Boundary[\CommitteeGraph{\Committee}]{\VIHPrefix{t}}}
=\sum_{x=1}^{t}\max\braces{0,\VIHRightReach{\Committee}{x}-t}.
\tag{9}
\]
This identity counts simple graph edges exactly: even if several selected candidates generate the same edge, the edge appears only once, and the rightmost selected interval containing $x$ describes the entire contiguous set of rightward neighbors of $x$.

We call
\[
\VIHRightReach{\Committee}{x}-t
\]
the \emph{signed contribution} of voter $x$. Equivalently, voter \(x\) may be viewed as row \(x\) of the adjacency matrix; \(R_\Committee(x)-t\) measures how far that row extends beyond the prefix boundary, with negative values retained for algebraic bookkeeping. The signed contribution lower bounds the actual contribution of $x$ to the boundary:
\[
\max\braces{0,\VIHRightReach{\Committee}{x}-t} \ge \VIHRightReach{\Committee}{x}-t
\]
The signed contribution is only a lower-bound device; it need not equal the actual contribution of a voter. In particular, it may be negative. For the boundary we get
\[
\abs{\Boundary[\CommitteeGraph{\Committee}]{\VIHPrefix{t}}}
\ge\sum_{x=1}^{t}\VIHRightReach{\Committee}{x}-t.
\]

We now use these prefix-boundary calculations to show that every committee meeting the threshold must have the intended form.

For one complete neutral phase, the signed contributions of all $\VIHPhaseSpan$ voters in the phase sum to the same value $\VIHNeutralArea$, independently of which alternative $\VIHBranchA{p}{i}$ or $\VIHBranchB{p}{i}$ is selected for each item. Choosing $\VIHBranchA{p}{i}$ rather than $\VIHBranchB{p}{i}$ changes where the contribution appears within the phase, but not its total over the complete phase. Later, when considering a prefix that falls inside a phase, we use a window of the same length $\VIHPhaseSpan$ consisting of a suffix of the preceding phase and a prefix of the current one. For the monotone choices used by the intended committee—once an item switches from \(B\) to \(A\), it never switches back—the signed contribution of every such interior window is at least \(\VIHNeutralArea\). Since signed contribution lower-bounds the actual boundary, this gives a uniform lower bound on all nondiagnostic prefixes.

\subsection{Soundness- from VI to 3-PARTITION}
The soundness proof is composed of three parts:
\begin{itemize}
    \item Every feasible committee has \emph{normal form}: must select all "fixed" candidates and exactly one of $\VIHBranchA{p}{i},\VIHBranchB{p}{i}$ for each item in each phase.
    \item Every feasible committee admits monotonicity: $\VIHChoice{p}{i}=1\rightarrow\VIHChoice{p+1}{i}=1$ For every phase $p$.
    \item Every feasible committee admits the desired cumulative weight for each stage: the wight probes of stage $h$ derive $\VIHEncodedWeight{2h}\ge hB$ and $\VIHEncodedWeight{2h+1}\le hB$. 
\end{itemize}

\paragraph{Normal Form}

We first derive the following lemma.

\begin{lemma}[Deletion lemma]
\label{lm:deletion-lemma}
Let $K$ be a committee, let $c\in K$, and let $u,v\in K\setminus\{c\}$
be respectively the selected candidates immediately preceding and following
$c$ in the interval order. Write their supports as
\[
[\ell_x+1,r_x],
\qquad x\in\{u,c,v\}.
\]
For a prefix $P_t=\{1,\ldots,t\}$, deleting $c$ from the committee removes
exactly
\[
\bigl[\min\{t,\ell_v\}-\ell_c\bigr]_+
\bigl[r_c-\max\{t,r_u\}\bigr]_+
\]
edges from $\partial(P_t)$, where $[x]_+=\max\{x,0\}$.
\end{lemma}

\begin{proof}
A voter can lose neighbors when $c$ is deleted only if $c$ is responsible
for extending its rightmost reach beyond that provided by the preceding
selected candidate $u$. These are precisely the voters
\[
\ell_c < x\le \ell_v.
\]
Among them, only voters lying in $P_t$ contribute to the loss across the
prefix cut, so the number of affected rows is
\[
\bigl[\min\{t,\ell_v\}-\ell_c\bigr]_+.
\]

For every such affected row, candidate $c$ extends the rightmost reach to
$r_c$, whereas after deleting $c$ the reach falls to $r_u$. Only neighbors
lying to the right of the cut contribute to $\partial(P_t)$, so each
affected row loses
\[
\bigl[r_c-\max\{t,r_u\}\bigr]_+
\]
crossing edges.

Multiplying the number of affected rows by the loss per row gives
\[
\bigl[\min\{t,\ell_v\}-\ell_c\bigr]_+
\bigl[r_c-\max\{t,r_u\}\bigr]_+,
\]
as claimed.
\end{proof}

\paragraph{Calculation of $\VIHBoundaryTarget$.}
We first calculate the signed contribution of a complete neutral phase and use it to express the target boundary threshold.
Fix a neutral phase $p$, and measure voter positions locally from
$\VIHPhaseOrigin{p}$.  For $1\le x\le \VIHPhaseSpan$, let
$\VIHNeutralRow{p}{x}$ denote the signed contribution of voter
$\VIHPhaseOrigin{p}+x$ to the prefix ending at
$\VIHPhaseOrigin{p+1}$, and let
\[
\VIHNeutralPrimitive{p}{u}
    :=\sum_{x=1}^{u}\VIHNeutralRow{p}{x}.
\]
Thus $\VIHNeutralPrimitive{p}{\VIHPhaseSpan}$ is the total signed
contribution of the rows of phase $p$ across its right boundary.

Recall that a candidate represented by the local pair $(a,b)$ begins at
local voter $a+1$ and has right endpoint
$\VIHPhaseOrigin{p+1}+b$.  Since the candidate intervals in the neutral
template are proper and consecutive candidates overlap, between two
consecutive candidate starts the rightmost reach is given by the candidate
that started most recently.  Hence, if a candidate with pair $(a,b)$ is
followed by a candidate starting at $a'+1$, then it contributes
\[
(a'-a)b
\]
to $\VIHNeutralPrimitive{p}{\VIHPhaseSpan}$.

We first observe that the total does not depend on which alternative is
chosen for an item.  Fix item $i$ and abbreviate
$\xi=\VIHItemOrigin{i}$ and $v=\VIHItemScale$.  Immediately before the
alternative pair, the rightmost reach has offset $\xi-v$.  If
$\VIHBranchA{p}{i}$ is selected instead of $\VIHBranchB{p}{i}$, then on
the single row at local position $\xi+v$ the signed contribution increases
by
\[
(\xi+v)-(\xi-v)=2v.
\]
On the following $2v$ rows, from $\xi+v+1$ through $\xi+3v$, the right
endpoint of $\VIHBranchA{p}{i}$ is one position earlier than that of
$\VIHBranchB{p}{i}$, decreasing the signed contribution by $1$ on each
row.  The total change is therefore
\[
2v-2v=0.
\]
Consequently,
\[
\VIHNeutralPrimitive{p}{\VIHPhaseSpan}
\]
is independent of all alternative choices.  We may therefore calculate it
assuming that $\VIHBranchB{p}{i}$ is selected for every item $i$.

Under this choice, the scaffold candidates control successive row blocks
of lengths
\[
H,\quad 1,\quad 2H-1,\quad H,\quad 1,\quad 4H-1,\quad v,
\]
with respective signed contributions per row
\[
-2H,\quad -H,\quad 0,\quad 5H,\quad 6H,\quad 7H,\quad 8H-v.
\]
Their total contribution is therefore
\begin{align*}
L_{\mathrm{sc}}
&=
H(-2H)
+1(-H)
+(2H-1)0
+H(5H)
+1(6H) \\
&\qquad
+(4H-1)(7H)
+v(8H-v)\\
&=
31H^2+8Hv-v^2-2H.
\end{align*}

Now consider item $i$.  Write $\xi_i=\VIHItemOrigin{i}$.  Apart from the
extra terminal part after the last item, the three relevant row blocks are
controlled by $\VIHBranchB{p}{i}$,
$\VIHMergeCandidate{p}{i}$, and $\VIHItemAnchor{p}{i}$.
Their lengths are respectively
\[
2v,\qquad v,\qquad v,
\]
and their signed contributions per row are
\[
\xi_i+v+1,\qquad
\xi_i+v+1+J,\qquad
\xi_i+3v.
\]
Thus the standard contribution of item $i$ is
\begin{align*}
L_i
&=
2v(\xi_i+v+1)
+v(\xi_i+v+1+J)
+v(\xi_i+3v)\\
&=
4v\xi_i+6v^2+v(J+3).
\end{align*}

Since
\[
\xi_i=8H+4v(i-1),
\]
we have
\[
\sum_{i=1}^{q}\xi_i
=
8qH+2q(q-1)v,
\]
and hence
\begin{align*}
\sum_{i=1}^{q}L_i
&=
32qHv
+(8q^2-2q)v^2
+qv(J+3).
\end{align*}

For the last item, $g_q$ remains the rightmost-started candidate for
$4H$ rows, whereas the standard item calculation above counted only $v$
of them.  We must therefore add the terminal contribution
\begin{align*}
L_{\mathrm{end}}
&=
(4H-v)(\xi_q+3v)\\
&=
(4H-v)\bigl(8H+(4q-1)v\bigr)\\
&=
32H^2+(16q-12)Hv-(4q-1)v^2.
\end{align*}

Combining the scaffold, item, and terminal contributions gives
\begin{align*}
\VIHNeutralPrimitive{p}{\VIHPhaseSpan}
&=
L_{\mathrm{sc}}
+\sum_{i=1}^{q}L_i
+L_{\mathrm{end}}\\
&=
63H^2
+(48q-4)Hv
+(8q^2-6q)v^2
+qv(J+3)
-2H\\
&=
\VIHNeutralArea.
\end{align*}

Thus every complete neutral phase has the same total signed contribution,
independently of its alternative choices:
\[
\VIHNeutralPrimitive{p}{\VIHPhaseSpan}
=
\VIHNeutralArea.
\]

We set $\VIHBoundaryTarget=\VIHNeutralArea-\VIHSlack$.

Finally, We use Lemma~\ref{lm:deletion-lemma} to show that every feasible committee has normal form.

\begin{lemma}[Normal form]
Let $K$ be a committee of size $k$ such that
\[
h(G_K)\ge \rho.
\]
Then $K$ contains every fixed candidate and, for every phase--item pair
$(p,i)$, exactly one of the two alternative candidates
$\VIHBranchA{p}{i}$ and $\VIHBranchB{p}{i}$.
\end{lemma}

\begin{proof}
As previously argued, every diagnostic prefix in the active region
must have boundary at least $T$.

We first establish a uniform reserve bound. At every diagnostic prefix used
below, even the graph obtained by selecting all available candidates has
boundary at most
\[
T+R,
\qquad
R:=W+\nu\Lambda.
\]
Hence, if omitting a candidate $c$ decreases the boundary of some diagnostic
prefix by more than $R$, then $c$ must belong to every feasible committee.
Indeed, if the deletion loss at $P_t$ is $\Delta_c>R$, then any committee
omitting $c$ satisfies
\[
|\partial_{G_K}(P_t)|
\le T+R-\Delta_c
<T,
\]
contradicting feasibility.

We now apply the deletion-rectangle lemma to the fixed candidates. For each
fixed candidate, we choose the corresponding diagnostic prefix and obtain the
following deletion losses:
\[
2H^2,\qquad
H-\delta_p,\qquad
(4H-1)(H+\delta_p),
\]
\[
(v-1)(H-v),\qquad
(v+\sigma_p a_i)J,\qquad
(v-1)(2v-1-J),
\]
together with the analogous losses for the remaining scaffold candidates.
By the parameter inequalities, every relevant deletion loss is larger than
$R$. Therefore every such fixed candidate must be selected.

The endpoint candidates are forced directly. If the head candidate is absent,
voter $1$ is isolated, while if the tail candidate is absent, voter $n$ is
isolated. In either case,
\[
h(G_K)=0<\rho.
\]
Thus every feasible committee contains all fixed candidates.

There are
\[
r(2q+7)+6
\]
fixed candidates.

It remains to consider the alternative candidates. Fix a phase--item pair
$(p,i)$ and suppose that neither
$\VIHBranchA{p}{i}$ nor $\VIHBranchB{p}{i}$ is selected.
At the corresponding forcing prefix, once
$\VIHBranchA{p}{i}$ is absent, deleting
$\VIHBranchB{p}{i}$ causes a loss of
\[
(2v-\sigma_p a_i)(2v+1).
\]
By the parameter inequalities,
\[
(2v-\sigma_p a_i)(2v+1)>R.
\]
Hence the boundary of that prefix would be smaller than $T$, contradicting
feasibility. Therefore every feasible committee contains at least one of
$\VIHBranchA{p}{i}$ and $\VIHBranchB{p}{i}$ for every $(p,i)$.

There are $rq$ such alternative pairs. Consequently, every feasible committee
contains at least
\[
r(2q+7)+6+rq
=
r(3q+7)+6
=
k
\]
candidates. Since the committee size is exactly $k$, all of these lower bounds
must be tight. Hence every fixed candidate is selected and exactly one
alternative candidate is selected from every pair
$\{\VIHBranchA{p}{i},\VIHBranchB{p}{i}\}$.
\end{proof}

\paragraph{Monotonicity.}

By the normal-form lemma, every feasible committee contains exactly one of
$\VIHBranchA{p}{i}$ and $\VIHBranchB{p}{i}$ for every phase--item pair
$(p,i)$.  We may therefore define
\[
z_{p,i}
:=
\begin{cases}
1, & \text{if $\VIHBranchA{p}{i}$ is selected},\\
0, & \text{if $\VIHBranchB{p}{i}$ is selected}.
\end{cases}
\]

We show that these choices are monotone across phases.  For every $p\ge 1$
and item $i$, consider the marker prefix ending at
\[
\mu_{p,i}
:=
\VIHPhaseOrigin{p}+\VIHItemOrigin{i}+\VIHItemScale,
\]
the leftmost endpoint of the support of candidate $\VIHBranchA{p}{i}$. At this prefix, the boundary satisfies
\[
|\partial(P_{\mu_{p,i}})|
=
\VIHNeutralArea
+\Lambda z_{p,i}
-\Lambda z_{p-1,i}
+C_{p,i},
\]
where $C_{p,i}$ collects the phase-specific perturbations.  These
perturbations satisfy $|C_{p,i}|\le \Xi$.

Indeed, up to the neutral baseline $\VIHNeutralArea$, changing the choice in
phase $p$ from $\VIHBranchB{p}{i}$ to $\VIHBranchA{p}{i}$ increases the
boundary at the marker by $\Lambda$, whereas making the same change in the
preceding phase decreases it by $\Lambda$.  All remaining phase-specific
changes contribute the error term $C_{p,i}$.

Suppose now that, for some $p$ and $i$,
\[
z_{p-1,i}=1
\qquad\text{and}\qquad
z_{p,i}=0.
\]
Then
\[
|\partial(P_{\mu_{p,i}})|
=
\VIHNeutralArea-\Lambda+C_{p,i}
\le
\VIHNeutralArea-\Lambda+\Xi.
\]
By the parameter inequalities, $\Lambda>\varepsilon+\Xi$.
Since $T=\VIHNeutralArea-\varepsilon$,
we get that
\[
|\partial(P_{\mu_{p,i}})|
<
\VIHNeutralArea-\varepsilon
=
T.
\]
This contradicts the common-threshold condition for feasible committees.

\paragraph{Cumulative Weight.}

For every phase $p$, define the cumulative weight encoded by that phase as
\[
w_p:=\sum_{i=1}^{q} a_i z_{p,i}.
\]
Recall that $\sigma_p=(-1)^p$.  We now show that the designated weight
probe in phase $p$ compares $w_p$ with the intended cumulative target
$\lfloor p/2\rfloor B$.

For every variable phase $p$, let
\[
\tau_p:=X_{p+1}+3H.
\]
We claim that
\[
\boxed{
|\partial(P_{\tau_p})|-T
=
\sigma_p
\left(
w_p-\left\lfloor\frac p2\right\rfloor B
\right).
}
\]

We obtain this identity by comparing the boundary at $\tau_p$ with the
neutral baseline $L$.  Fix an item $i$.  The displacement of the weighting
candidate $e_{p,i}$ from its neutral position changes the boundary at the
probe by
\[
\sigma_p J a_i.
\]
This contribution is independent of which alternative is selected.

The choice between $A_{p,i}$ and $B_{p,i}$ contributes an additional
$\sigma_p a_i$ precisely when $A_{p,i}$ is selected.  To see this, compare
the two choices at the probe.  The unique row at the left endpoint of
$A_{p,i}$ gains $\Lambda=2v$ crossing edges when $A_{p,i}$ is selected.
On the following
\[
\Lambda-\sigma_p a_i
\]
rows, the rightmost reach is one position smaller than under the
$B_{p,i}$ choice.  Hence the net difference is
\[
\Lambda-(\Lambda-\sigma_p a_i)
=
\sigma_p a_i.
\]
Since this term occurs exactly when $z_{p,i}=1$, the total contribution of
item $i$ relative to the neutral phase is
\[
\sigma_p J a_i+\sigma_p a_i z_{p,i}.
\]

Summing over all items and using
\[
\sum_{i=1}^{q}a_i=W
\]
gives
\[
\sigma_p JW+\sigma_p w_p
=
\sigma_p(JW+w_p).
\]

The calibration candidates contribute the remaining phase-dependent term.
At the probe $\tau_p$, their total change from the neutral baseline is
\[
\delta_{p+1}-\delta_p.
\]
Consequently,
\[
|\partial(P_{\tau_p})|
=
L+\sigma_p(JW+w_p)+\delta_{p+1}-\delta_p.
\]

By the definition of the calibration offsets,
\[
\delta_{p+1}-\delta_p
=
-\varepsilon
-\sigma_p
\left(
JW+\left\lfloor\frac p2\right\rfloor B
\right).
\]
Substituting this into the preceding expression yields
\begin{align*}
|\partial(P_{\tau_p})|
&=
L+\sigma_p(JW+w_p)
-\varepsilon
-\sigma_p
\left(
JW+\left\lfloor\frac p2\right\rfloor B
\right)\\
&=
L-\varepsilon
+\sigma_p
\left(
w_p-\left\lfloor\frac p2\right\rfloor B
\right).
\end{align*}
Since $T=L-\varepsilon$, we obtain
\[
\boxed{
|\partial(P_{\tau_p})|-T
=
\sigma_p
\left(
w_p-\left\lfloor\frac p2\right\rfloor B
\right).
}
\]

Now let $K$ be a feasible committee.  By the common-threshold lemma,
\[
|\partial(P_{\tau_p})|\ge T
\]
for every weight probe.  Hence
\[
\sigma_p
\left(
w_p-\left\lfloor\frac p2\right\rfloor B
\right)
\ge 0.
\]

For every $0\le h<s$, the even phase $p=2h$ therefore gives
\[
w_{2h}\ge hB,
\]
whereas the following odd phase $p=2h+1$ gives
\[
w_{2h+1}\le hB.
\]
By monotonicity,
\[
z_{2h,i}\le z_{2h+1,i}
\qquad\text{for every }i,
\]
and since all source weights are positive,
\[
w_{2h}\le w_{2h+1}.
\]
Thus
\[
hB
\le
w_{2h}
\le
w_{2h+1}
\le
hB,
\]
and consequently
\[
\boxed{
w_{2h}=w_{2h+1}=hB
}
\qquad
\text{for every }0\le h<s.
\]

For the final even phase $p=2s$, the weight probe gives
\[
w_{2s}\ge sB.
\]
On the other hand,
\[
w_{2s}
\le
\sum_{i=1}^{q}a_i
=
W
=
sB.
\]
Therefore
\[
\boxed{w_{2s}=sB.}
\]

The sets encoded by the even phases therefore have cumulative weights
$0,B,2B,\ldots,sB$.  By monotonicity, the differences between consecutive
stages form disjoint bins, and each such difference has total weight exactly
$B$.  As noted in the reduction idea, the strict bounds
\[
\frac{B}{4}<a_i<\frac{B}{2}
\]
imply that every bin contains exactly three items.  Hence these stage
differences form a valid solution to the source
\textsc{Strict-Unary-3-Partition} instance.

\subsubsection{Completeness}

Suppose now that the source instance admits a valid partition into bins
$D_0,\ldots,D_{s-1}$, each of total weight $B$.  For each item $i$, let
$\operatorname{bin}(i)$ denote the index of its bin.  We select every fixed
candidate and, in phase $p$, select
\[
\VIHBranchA{p}{i}
\quad\Longleftrightarrow\quad
\operatorname{bin}(i)
<
\left\lfloor\frac p2\right\rfloor,
\]
and select $\VIHBranchB{p}{i}$ otherwise.

This committee has exactly the required size.  Its alternative choices are
monotone across the phases, and the cumulative weight encoded by phase $p$ is
\[
w_p
=
\left\lfloor\frac p2\right\rfloor B.
\]
Consequently, by the weight-probe identity proved above, every weight probe
has boundary exactly $T$.

It remains to show that no other cut has expansion below the target.

\paragraph{Prefix cuts.}

We first consider prefixes of the voter order.  Temporarily return to the
neutral construction: set all calibration offsets to zero and move each
weighting candidate $e_{p,i}$ back to its neutral starting position
$\xi_i+3v$.

For a neutral phase $p$, let $F_p(u)$ denote the total signed contribution
of its first $u$ rows.  We have already shown that every complete neutral
phase satisfies
\[
F_p(\VIHPhaseSpan)=\VIHNeutralArea.
\]

Changing the choice of a single item from $B$ to $A$ adds a nonnegative
quantity to $F_p(u)$ at every intermediate position $u$: the difference
starts at zero, rises when the left endpoint of $A$ is reached, and then
decreases back to zero before the end of the item pattern.  Since the
committee constructed above makes monotone choices across phases, it follows
that
\[
F_{p-1}(u)\le F_p(u)
\qquad
\text{for every }0\le u\le\VIHPhaseSpan.
\]

Consider a prefix whose endpoint lies $u$ positions into phase $p$.  The
suffix of phase $p-1$ together with the prefix of phase $p$ contains exactly
$\VIHPhaseSpan$ rows, and its signed contribution is
\begin{align*}
F_{p-1}(\VIHPhaseSpan)
-F_{p-1}(u)
+F_p(u)
&\ge
F_{p-1}(\VIHPhaseSpan)\\
&=
\VIHNeutralArea.
\end{align*}
The actual contribution of every row is at least its signed contribution,
and rows outside this window contribute a nonnegative number of crossing
edges.  Hence every interior prefix of the neutral construction has boundary
at least $\VIHNeutralArea$.

The left and right ends require only the corresponding endpoint check.
At the left end, define
\[
[u]_{[a,b]}
:=
\max\{0,\min\{u,b\}-a\}.
\]
A direct row count shows that the head candidate together with the truncated
scaffold of phase $0$ changes the hypothetical complete-neutral profile by
\[
\Gamma(u)
=
2H[u]_{[0,H]}
+
H[u]_{[H,H+1]}
\ge0.
\]
Thus the missing scaffold candidates at the beginning do not decrease the
neutral lower bound.  At the other end, the final scaffold-only phase
provides exactly the continuation of the row-reach profile required by the
same argument up to $\VIHActiveEnd$; the tail begins only after this point.
Therefore
\[
\boxed{
|\partial_{\mathrm{neutral}}(P_t)|
\ge
\VIHNeutralArea
\qquad
\text{for every }
\VIHPhaseOrigin{1}\le t\le\VIHActiveEnd.
}
\]

We now restore the phase-specific modifications.  The calibration offsets can
affect a prefix only near a scaffold transition.  Outside these neighborhoods,
the two opposite calibration shifts cancel exactly.  Inside such a
neighborhood, but away from the weight probe, direct counting of the scaffold
rows gives the stronger neutral bound
\[
|\partial_{\mathrm{neutral}}(P_t)|
\ge
\VIHNeutralArea+H.
\]
Since $|\delta_p|\le D$, restoring the calibration can decrease the boundary
there by at most $2D$.

Finally, restore the source-weight displacement of the candidates $e_{p,i}$.
Moving the start of $e_{p,i}$ by $a_i$ positions can change the boundary of
any voter set by at most
\[
a_i(J+1).
\]
Since the total source weight is $W$, the change contributed by one phase is
at most
\[
\Xi=(J+1)W,
\]
and over all variable phases the total possible change is at most $r\Xi$.

We can therefore distinguish three kinds of active prefixes.  Outside the
calibration neighborhoods,
\[
|\partial(P_t)|
\ge
\VIHNeutralArea-r\Xi
>
\VIHNeutralArea-\varepsilon
=
T.
\]
Inside a calibration neighborhood but away from its weight probe,
\[
|\partial(P_t)|
\ge
\VIHNeutralArea+H-2D-r\Xi
\ge
T,
\]
by the parameter inequalities.  At the weight probe itself we already know
that
\[
|\partial(P_{\tau_p})|=T.
\]
Hence
\[
\boxed{
|\partial(P_t)|\ge T
\qquad
\text{for every }
\VIHPhaseOrigin{1}\le t\le\VIHActiveEnd.
}
\]

\paragraph{Arbitrary cuts.}

The preceding argument only concerns prefixes, whereas the Cheeger constant
minimizes over all voter sets.  We now show that the same threshold holds for
every cut of the constructed committee graph.

Put
\[
a:=\VIHPhaseOrigin{1},
\qquad
b:=\VIHActiveEnd,
\]
and define
\[
K_L:=P_a,
\qquad
W_{\mathrm{mid}}:=P_b\setminus P_a,
\qquad
K_R:=V\setminus P_b.
\]
The head and tail candidates make $K_L$ and $K_R$ cliques.  The parameter
choices give
\[
b-a\le Q,
\qquad
b\le QT,
\qquad
n\le2Q(T+1),
\]
and
\[
T+2\le a,
\qquad
T+2\le n-b,
\qquad
n<2(n-b).
\]
In particular, both endpoint cliques have size at least $T+2$, while
$K_R$ contains more than half of the voters.

The graph induced by the constructed committee is connected: the head,
the scaffold candidates $c_3$ in successive phases, and the tail form a
chain of overlapping selected intervals.

We first establish one consequence of the voter-interval structure.  If
$i<j<k$ and voters $i$ and $k$ are adjacent, then the selected candidate
that they jointly approve is approved by every voter between them.  Hence
\[
i<j<k,\qquad i\sim k
\quad\Longrightarrow\quad
i\sim j
\ \text{ and }\
j\sim k.
\]

Orient every graph edge from its smaller endpoint to its larger endpoint.
For a voter set $U$, let
\[
F^+(U)
:=
\bigl|
\{\,ij\in E:
i<j,\ i\in U,\ j\notin U\,\}
\bigr|
\]
be the number of crossing edges directed from $U$ to its complement.
Clearly
\[
F^+(U)\le|\partial(U)|,
\]
and equality holds whenever $U$ is a prefix.

\begin{lemma}[Separator lemma]
If
\[
K_L\subseteq U\subseteq V\setminus K_R,
\]
then
\[
|\partial(U)|\ge T.
\]
\end{lemma}

\begin{proof}
Minimize $F^+(U)$ over all sets satisfying
\[
K_L\subseteq U\subseteq V\setminus K_R,
\]
and, among all minimizers, choose one of maximum cardinality.

Suppose that its membership vector contains consecutive entries $(0,1)$ at
voters $j,j+1$.  Both voters lie in $W_{\mathrm{mid}}$, so either one may
change sides without violating the endpoint conditions.

Let
\[
\alpha
=
|\{i<j:i\in U,\ i\sim j\}|,
\qquad
\beta
=
|\{i<j:i\in U,\ i\sim j+1\}|,
\]
and let
\[
\gamma
=
|\{i>j+1:i\notin U,\ j\sim i\}|,
\qquad
\delta
=
|\{i>j+1:i\notin U,\ j+1\sim i\}|.
\]
By the interval property above, every earlier neighbor of $j+1$ is also a
neighbor of $j$, while every later neighbor of $j$ is also a neighbor of
$j+1$.  Therefore
\[
\alpha\ge\beta,
\qquad
\delta\ge\gamma.
\]

If we add $j$ to $U$, then
\[
F^+(U\cup\{j\})-F^+(U)
=
\gamma-\alpha.
\]
If $\gamma\le\alpha$, this operation does not increase $F^+$.  It either
decreases it, contradicting minimality, or preserves it while increasing
$|U|$, contradicting the choice of the minimizer.

We may therefore assume $\gamma>\alpha$.  Instead remove $j+1$ from $U$.
Then
\[
F^+(U\setminus\{j+1\})-F^+(U)
=
\beta-\delta
\le
\alpha-\gamma
<
0,
\]
again contradicting minimality.

Thus a minimizing set has no adjacent $(0,1)$ transition.  Since it contains
the entire left endpoint clique and excludes the entire right endpoint
clique, it must be a prefix $P_t$ for some
\[
a\le t\le b.
\]
The prefix bound proved above gives
\[
F^+(P_t)
=
|\partial(P_t)|
\ge T.
\]
Hence the minimum possible value of $F^+$ under the endpoint conditions is
at least $T$, and consequently every admissible $U$ satisfies
\[
|\partial(U)|
\ge
F^+(U)
\ge
T.
\qedhere
\]
\end{proof}

We can now consider an arbitrary nonempty voter set $U$ with
\[
|U|\le\frac n2.
\]

If $U$ splits either endpoint clique, let that clique have size $d$ and let
$s'$ of its voters lie in $U$, where $1\le s'<d$.  The edges internal to
the clique already contribute
\[
s'(d-s')
\ge
d-1
\ge
T+1
\]
crossing edges.  Hence
\[
|U|
\le
\frac n2
\le
Q(T+1)
\le
Q|\partial(U)|.
\]

Suppose next that neither endpoint clique is split and
\[
U\cap K_L=\varnothing.
\]
The set $U$ cannot contain $K_R$, since $|K_R|>n/2$.  Therefore
\[
U\subseteq W_{\mathrm{mid}},
\]
and hence
\[
|U|
\le
|W_{\mathrm{mid}}|
=
b-a
\le
Q.
\]
Since the graph is connected and $U$ is a nonempty proper subset,
\[
|\partial(U)|\ge1,
\]
so again
\[
|U|\le Q|\partial(U)|.
\]

The only remaining possibility is that neither endpoint clique is split and
\[
K_L\subseteq U.
\]
Again $U$ cannot contain $K_R$, so the separator lemma applies and gives
\[
|\partial(U)|\ge T.
\]
Moreover,
\[
|U|\le b\le QT.
\]
Thus
\[
|U|\le Q|\partial(U)|.
\]

These cases exhaust all nonempty sets with $|U|\le n/2$.  Therefore
\[
\frac{|\partial(U)|}{|U|}
\ge
\frac1Q
=
\rho
\]
for every such $U$, and hence
\[
\boxed{
h(G_K)\ge\rho.
}
\]
This proves completeness.

\subsubsection{Polynomiality}

It remains to verify that the construction has polynomial size.  Let $\ell$
denote the unary encoding length of the source instance.  With the source
notation used above,
\[
\ell=(s+1)B+3s+1,
\]
so in particular
\[
s,\ B,\ W\le\ell.
\]
Substituting these bounds successively into the parameter definitions shows
that the number of voters satisfies
\[
n=O(\ell^{19}),
\]
while the number of candidates is
\[
O(\ell^2).
\]
Thus even an explicit voter--candidate approval matrix contains only
\[
O(\ell^{21})
\]
entries.

All construction parameters and interval endpoints are obtained from the
source instance by a fixed number of additions and multiplications of
polynomially bounded integers.  The reduction can therefore generate the
voters and candidates and determine every approval by testing membership in
the corresponding voter interval, all in time polynomial in $\ell$.

\subsection{Type-graph structure and exact evaluation}
\label{app:type-graph-proof}

For completeness, we give the structural proof underlying
Theorem~\ref{thm:max_single_split_node} and the exact type-graph search used
in Corollary~\ref{cor:type_graph_cheeger_evaluation}.

Fix a committee graph $G=(V,E)$.  When the committee is fixed, write
$\vtype{v}$ for a voter's committee-relative type,
$\TypeSet{V}$ for the occurring types,
$\TypeClass{\alpha}$ for the corresponding voter class, and
$\TypeWeight{\alpha}$ for its size.

\begin{lemma}[Completion of a partially selected type]
\label{lm:type_all_or_nothing}
Let
\[
    c:=\left\lfloor\frac{\abs V}{2}\right\rfloor,
\]
and let $S\subseteq V$ be a Cheeger-optimal smaller side with
$1\le\abs S<c$.  Suppose that $S$ splits a type $\alpha$.  Then, for
every
\[
    U\subseteq\TypeClass{\alpha}\setminus S
    \qquad\text{with}\qquad
    \abs U\le c-\abs S,
\]
the set $S\cup U$ is also Cheeger-optimal.  Consequently $S$ can be
extended, without losing optimality, until either the whole type is selected
or the cut has cardinality $c$.
\end{lemma}

\begin{proof}
Write $h:=\Cheeger{G}$ and choose
$v\in S\cap\TypeClass{\alpha}$.  Put $R:=S\setminus\{v\}$.  Since $S$ is
optimal,
\[
    \abs{\Boundary[G]{S}}=h\abs S,
\]
whereas
\[
    \abs{\Boundary[G]{R}}\ge h\abs R.
\]
For $R\ne\emptyset$ this is the definition of $h$, because
$\abs R<c$; for $R=\emptyset$ it is $0\ge0$.  Hence
\begin{equation}
\label{eq:first_same_type_marginal}
    \abs{\Boundary[G]{S}}-\abs{\Boundary[G]{R}}\le h.
\end{equation}

All voters in $\TypeClass{\alpha}$ have the same neighbors outside the
type class.  If $\alpha\ne\emptyset$, the type class is a clique, so after
one voter of this type is added there is one fewer same-type vertex outside
the cut and one more inside; the next boundary marginal is therefore smaller
by $2$.  If $\alpha=\emptyset$, there are no same-type edges and the marginal
is unchanged.  Thus successive additions of voters of type $\alpha$ have
nonincreasing boundary marginals.

Every voter added from $\TypeClass{\alpha}\setminus S$ therefore has
boundary marginal at most the quantity in
Equation~\eqref{eq:first_same_type_marginal}, and hence at most $h$.  For
every admissible $U$,
\[
    \abs{\Boundary[G]{S\cup U}}
    \le h\abs S+h\abs U
    =h\abs{S\cup U}.
\]
Because $1\le\abs{S\cup U}\le c$, the definition of $h$ gives the reverse
inequality.  Equality follows, so $S\cup U$ is Cheeger-optimal.
\end{proof}

\begin{proof}[Proof of Theorem~\ref{thm:max_single_split_node}]
Start from any Cheeger-optimal smaller side $S$, with
$1\le\abs S\le c$.  While $\abs S<c$ and $S$ splits a type $\alpha$,
apply Lemma~\ref{lm:type_all_or_nothing} to add voters of type $\alpha$
until either $\TypeClass{\alpha}\subseteq S$ or $\abs S=c$.  Every such
extension preserves optimality.  If a type is completed below the cap, the
number of split types decreases.  Thus if the process terminates below the
cap, the resulting optimum is type-respecting.

Now suppose $\abs S=c$ and $S$ splits two distinct types.  Choose
\[
    u,v\in S,
    \qquad
    u'\in\TypeClass{\vtype{u}}\setminus S,
    \qquad
    v'\in\TypeClass{\vtype{v}}\setminus S,
\]
with $\vtype{u}\ne\vtype{v}$, and consider
\[
    S_u=S\setminus\{u\}\cup\{v'\},
    \qquad
    S_v=S\setminus\{v\}\cup\{u'\}.
\]
Set
\[
    \Delta_u
    :=
    \abs{\Boundary[G]{S_u}}-\abs{\Boundary[G]{S}},
    \qquad
    \Delta_v
    :=
    \abs{\Boundary[G]{S_v}}-\abs{\Boundary[G]{S}}.
\]
Optimality and size preservation give $\Delta_u,\Delta_v\ge0$.

Voters of the same type have identical neighborhoods outside their own type
class.  Adding the two swap differences therefore cancels all contributions
from vertices outside the two involved classes, leaving
\begin{align}
\label{eq:two_type_swap_sum}
    \Delta_u+\Delta_v
    ={}&
    4\,\mathbf 1[\vtype{u}\cap\vtype{v}\ne\emptyset]
    -2\,\mathbf 1[\vtype{u}\ne\emptyset]
    -2\,\mathbf 1[\vtype{v}\ne\emptyset]
    \le0.
\end{align}
If the two types intersect, both are nonempty and the displayed quantity is
zero; if they are disjoint, the first term vanishes.  Since the two types are
distinct, they cannot both be empty.  Together with
$\Delta_u,\Delta_v\ge0$, this forces both swap costs to be zero.

Repeat the zero-cost transfer, removing one selected voter of the first type
and adding one unselected voter of the second type.  At every intermediate
cut both types are still split until one reaches an endpoint; the same
two-swap calculation applies because the cut remains cap-size and optimal.
After finitely many steps one of the two types is no longer split and no
other type membership has changed.  Repeating whenever two split types
remain terminates with an optimal cap-size side that splits at most one type.
This proves the theorem.
\end{proof}

\paragraph{Exact evaluation from the type graph.}
Let
\[
    T:=\TypeSet{V},
    \qquad
    n_\alpha:=\TypeWeight{\alpha},
    \qquad
    c:=\left\lfloor\frac{\abs V}{2}\right\rfloor.
\]
For $\mathcal X\subseteq T$, put
\[
    s_{\mathcal X}
    :=
    \sum_{\beta\in\mathcal X}n_\beta
\]
and
\[
    B_{\mathcal X}
    :=
    \sum_{\substack{
        \beta\in\mathcal X,\ \gamma\in T\setminus\mathcal X\\
        \{\beta,\gamma\}\in\TypeGraphEdges{G}
    }}
    n_\beta n_\gamma.
\]
This is exactly the boundary of the type-respecting side represented by
$\mathcal X$.

For the cap-size case, fix a prospective split type $\alpha\in T$ and a set
$\mathcal X\subseteq T\setminus\{\alpha\}$ of types lying wholly on the
selected side.  Set
\[
    x:=c-s_{\mathcal X}.
\]
Whenever $1\le x\le n_\alpha-1$, any choice of $x$ voters from
$\TypeClass{\alpha}$ yields the same boundary:
\begin{align}
\label{eq:split_type_boundary}
B_{\alpha,\mathcal X}(x)
={}&
\sum_{\substack{
    \beta\in\mathcal X,\
    \gamma\in T\setminus(\mathcal X\cup\{\alpha\})\\
    \{\beta,\gamma\}\in\TypeGraphEdges{G}
}}
n_\beta n_\gamma
\\
&+
\sum_{\substack{
    \beta\in\mathcal X\\
    \{\alpha,\beta\}\in\TypeGraphEdges{G}
}}
(n_\alpha-x)n_\beta
+
\sum_{\substack{
    \gamma\in T\setminus(\mathcal X\cup\{\alpha\})\\
    \{\alpha,\gamma\}\in\TypeGraphEdges{G}
}}
x n_\gamma
\notag\\
&+
\mathbf 1_{\alpha\ne\emptyset}\,x(n_\alpha-x).
\notag
\end{align}
The last term counts edges crossing inside a split nonempty type, which is a
clique.

Consequently
\begin{align}
\label{eq:exact_type_cheeger}
\Cheeger{G}
=
\min\Biggl\{
&
\min_{\substack{
    \mathcal X\subseteq T\\
    1\le s_{\mathcal X}\le c
}}
\frac{B_{\mathcal X}}{s_{\mathcal X}},
\\
&
\min_{\substack{
    \alpha\in T,\ \mathcal X\subseteq T\setminus\{\alpha\}\\
    x=c-s_{\mathcal X},\ 1\le x\le n_\alpha-1
}}
\frac{B_{\alpha,\mathcal X}(x)}{c}
\Biggr\}.
\notag
\end{align}
Every expression on the right is the expansion of an actual smaller side,
while Theorem~\ref{thm:max_single_split_node} guarantees that one of these
candidates is optimal.  The first minimum has at most $2^{\abs T}$ candidates
and the second at most $\abs T\,2^{\abs T-1}$ candidates, proving
Corollary~\ref{cor:type_graph_cheeger_evaluation}.

\end{document}